\documentclass[fleqn,usenatbib]{mnras}

\usepackage[T1]{fontenc}

\DeclareRobustCommand{\VAN}[3]{#2}
\let\VANthebibliography\thebibliography
\def\thebibliography{\DeclareRobustCommand{\VAN}[3]{##3}\VANthebibliography}

\usepackage{graphicx}	
\usepackage{amsmath}	

\usepackage{newtxtext,newtxmath}
\usepackage{pdflscape}

\newcommand\Msun{\,\rmn{M}_{\sun}}

\newcommand\Gyr{\,\rmn{Gyr}}
\newcommand\Myr{\,\rmn{Myr}}
\newcommand{\yr}{\,\rmn{yr}}
\newcommand{\pyr}{\,\rmn{yr}^{-1}}
\newcommand\pc{\,\rmn{pc}}

\newcommand\Gpc{\,\rmn{Gpc}}

\newcommand\Mag{\,\rmn{mag}}

\newcommand\FeH{\lbrack \rmn{Fe}/\rmn{H} \rbrack}

\newcommand\LogZ{\log(Z/\rmn{Z}_{\sun})}

\newcommand{\ts}{\textsuperscript}

\newcommand{\gadget}   {\textsc{gadget3}}
\newcommand{\subfind}  {\textsc{subfind}}

\newcommand{\cMpc}     {\,{\rm cMpc}}
\newcommand{\Mcstar}   {M_\rmn{c,\ast}}

\title[E-MOSAICS predictions for \textit{JWST} proto-GCs]{Comparing E-MOSAICS predictions of high-redshift proto-globular clusters with \textit{JWST} observations in lensed galaxies}

\author[J. Pfeffer et al.]{Joel Pfeffer,$^{1,2}$\thanks{E-mail: jpfeffer@swin.edu.au (JP)}
Duncan A.~Forbes,$^{1,2}$
Aaron J.~Romanowsky,$^{3,4}$
Nate Bastian,$^{5,6}$
Robert A.~Crain,$^{7}$
\newauthor
J.~M.~Diederik Kruijssen,$^{8,9}$
Kenji Bekki,$^{10}$
Jean P.~Brodie,$^{1,2,11}$
M\'{e}lanie Chevance,$^{12,9}$
Warrick J.~Couch,$^{1}$
\newauthor
Jonah S.~Gannon$^{1,2}$
\\
$^{1}$Centre for Astrophysics \& Supercomputing, Swinburne University, Hawthorn, VIC 3122, Australia\\
$^{2}$ARC Centre of Excellence for All Sky Astrophysics in 3 Dimensions (ASTRO 3D), Australia\\
$^{3}$Department of Physics and Astronomy, San Jos\'{e} State University, One Washington Square, San Jose, CA 95192, USA\\
$^{4}$Department of Astronomy \& Astrophysics, University of California Santa Cruz, 1156 High Street, Santa Cruz, CA 95064, USA\\
$^{5}$Donostia International Physics Center (DIPC), Paseo Manuel de Lardizabal, 4, E-20018 Donostia-San Sebasti\'{a}n, Guipuzkoa, Spain\\
$^{6}$IKERBASQUE, Basque Foundation for Science, E-48013 Bilbao, Spain\\
$^{7}$Astrophysics Research Institute, Liverpool John Moores University, 146 Brownlow Hill, Liverpool L3 5RF, UK\\
$^{8}$Technical University of Munich, School of Engineering and Design, Department of Aerospace and Geodesy, Chair of Remote Sensing Technology, \\\hspace{2.2mm}Arcisstr. 21, 80333 Munich, Germany\\
$^{9}$Cosmic Origins Of Life (COOL) Research DAO, coolresearch.io\\
$^{10}$International Centre for Radio Astronomy Research, University of Western Australia, 35 Stirling Highway, Crawley, WA 6009, Australia\\
$^{11}$University of California Observatories, 1156 High Street, Santa Cruz, CA 95064, USA\\
$^{12}$Zentrum f\"{u}r Astronomie der Universit\"{a}t Heidelberg, Institut f\"{u}r Theoretische Astrophysik, Albert-Ueberle-Str. 2, 69120 Heidelberg
}

\date{Accepted 2024 November 27. Received 2024 November 27; in original form 2024 October 09}

\pubyear{2024}

\begin{document}
\label{firstpage}
\pagerange{\pageref{firstpage}--\pageref{lastpage}}
\maketitle

\begin{abstract}
High-resolution imaging and strong gravitational lensing of high-redshift galaxies have enabled the detection of compact sources with properties similar to nearby massive star clusters.
Often found to be very young, these sources may be globular clusters detected in their earliest stages.
In this work, we compare predictions of high-redshift  ($z \sim 1$--$10$) star cluster properties from the E-MOSAICS simulation of galaxy and star cluster formation with those of the star cluster candidates in strongly lensed galaxies from \textit{James Webb} (\textit{JWST}) and \textit{Hubble Space Telescope} (\textit{HST}) imaging.
We select galaxies in the simulation that match the luminosities of the majority of lensed galaxies with star cluster candidates observed with \textit{JWST}.
We find that the luminosities, ages and masses of the brightest star cluster candidates in the high-redshift galaxies are consistent with the E-MOSAICS model.
In particular, the brightest cluster ages are in excellent agreement.
The results suggest that star clusters in both low- and high-redshift galaxies may form via common mechanisms.
However, the brightest clusters in the lensed galaxies tend to be $\approx 1$--$1.5 \Mag$ brighter and $\approx 0.5$~dex more massive than the median E-MOSAICS predictions.
We discuss the large number of effects that could explain the discrepancy, including simulation and observational limitations, stellar population models, cluster detection biases and nuclear star clusters.
Understanding these limitations would enable stronger tests of globular cluster formation models.
\end{abstract}




\section{Introduction}
\label{sec:introduction}

Star clusters are one of the most common types of stellar systems.
They range from the young star clusters (sometimes termed `young massive clusters' or `open clusters') found in star-forming galaxies \citep{Portegies-Zwart_McMillan_and_Gieles_10, Adamo_and_Bastian_18} to old globular clusters (GCs) found in nearly all galaxies with stellar masses $>10^9 \Msun$ \citep{Harris_91, Brodie_and_Strader_06}.
In galaxies like the Milky Way and M31, massive star clusters tend to be very old (i.e. the GCs, $\gtrsim 10^5 \Msun$, $\gtrsim 12 \Gyr$; \citealt{Forbes_and_Bridges_10, Caldwell_et_al_11, Dotter_Sarajedini_and_Anderson_11, VandenBerg_et_al_13, Usher_Caldwell_and_Cabrera-Ziri_24}), while low mass clusters tend to be very young \citep[e.g.][]{Johnson_et_al_16, Hunt_and_Reffert_24}, with only a few massive young clusters \citep{Gennaro_et_al_11, Davies_et_al_11}.
Other galaxies (such as M33 and the Magellanic Clouds) have formed massive clusters throughout their entire history \citep{Beasley_et_al_15, Horta_et_al_21a}.

While young star clusters have been found to be the densest regions of the hierarchical star formation process \citep{Longmore_et_al_14, Krumholz_McKee_and_Bland-Hawthorn_19}, the formation of old GCs has been widely debated due to the inability to observe their formation directly \citep{Kruijssen_14, Forbes_et_al_18}.
Scenarios for the formation of GCs broadly fall into two classes: special high-redshift conditions for GC formation \citep[which operates separately from the formation of young clusters or `intermediate age' GCs, e.g.][]{Peebles_84, Fall_and_Rees_85, Rosenblatt_Faber_and_Blumenthal_88, Naoz_and_Narayan_14, Trenti_Padoan_and_Jimenez_15, Mandelker_et_al_18, Creasey_et_al_19, Madau_et_al_20}; or, a common mechanism for both young and old star clusters, with the star formation conditions for massive cluster formation (e.g. higher pressures for star formation) generally being more prevalent in the high-redshift Universe \citep[e.g.][]{Ashman_and_Zepf_92, Harris_and_Pudritz_94, Kravtsov_and_Gnedin_05, Kruijssen_15, Li_et_al_17, P18, Ma_et_al_20, Horta_et_al_21a}.

The two classes of GC formation generally predict different formation epochs (e.g. $z \gtrsim 6$ for models of formation in dark matter minihaloes, \citealt{Trenti_Padoan_and_Jimenez_15, Boylan-Kolchin_17, Creasey_et_al_19, Valenzuela_et_al_21}; ages that follow host galaxy formation histories in `young cluster'-based models, \citealt{Muratov_and_Gnedin_10, Li_and_Gnedin_14, K19a}), thus the ages of GCs can (in principle) offer constraints on their formation process \citep[e.g. see][for a review]{Forbes_et_al_18}.
Unfortunately, even with resolved colour--magnitude diagram fitting of Milky Way GCs \citep[e.g.][]{Dotter_et_al_10, VandenBerg_et_al_13} the age uncertainties ($\sim 1 \Gyr$) of old stellar populations are too large to distinguish models.

One method that offers a window into the formation of GCs is strong gravitational lensing of high-redshift galaxies by foreground galaxy clusters, first enabled by observations with the \textit{Hubble Space Telescope} \citep[\textit{HST}, e.g.][]{Johnson_T_et_al_17, Vanzella_et_al_17a, Vanzella_et_al_17b, Vanzella_et_al_19}.
Such observations have now seen a drastic increase with the \textit{James Webb Space Telescope} (\textit{JWST}), with star cluster candidates (compact clumps) detected from redshifts $\sim 1$--$10$ \citep{Mowla_et_al_22, Mowla_et_al_24, Vanzella_et_al_22b, Vanzella_et_al_23, Claeyssens_et_al_23, Adamo_et_al_24, Fujimoto_et_al_24, Messa_et_al_24b}.
In ideal cases with magnifications of $\mu > 100$, lensing can provide resolution of $\approx 1 \pc$ (in the tangential direction of the arcs) with current instruments, enabling observations of star clusters in high-redshift galaxies and potentially catching GCs in their youngest stages \citep[e.g.][]{Vanzella_et_al_22a, Adamo_et_al_24}.
Tighter age constraints on the much younger stellar populations (typical ages $< 1 \Gyr$) may also enable stronger constraints on the epoch of GC formation.
These observations can be compared directly with predictions from GC formation models as strong tests of GC formation mechanisms.
Previous work based on \textit{HST} observations found that GC formation models \citep{Boylan-Kolchin_17, Pfeffer_et_al_19a} agree well with the UV luminosity function of compact sources (proto-GCs) at $z \sim 6$ \citep{Bouwens_et_al_21}.
However, there is yet to be a systematic comparison with star clusters in lensed galaxies across a wide redshift range.

In this work, we compare the observed properties of compact high-redshift clumps (star cluster candidates) determined with \textit{JWST} and \textit{HST} with predictions from the E-MOSAICS project \citep[MOdelling Star cluster population Assembly In Cosmological Simulations within EAGLE,][]{P18, K19a}.
In E-MOSAICS, both young and old star clusters are assumed to form and evolve following the same physical mechanisms.
In particular, star cluster formation is based on models which reproduce the scaling relations of young star cluster populations at $z=0$ \citep{Kruijssen_12, Reina-Campos_and_Kruijssen_17, Pfeffer_et_al_19b}.
We aim to test if these models can also explain the properties of GC candidates in high-redshift lensed galaxies.

This paper is structured as follows.
In Section~\ref{sec:methods} we describe the E-MOSAICS simulations, stellar population modelling, galaxy and star cluster selection, and the sample of lensed galaxies with star cluster candidates compiled from the literature.
In Section~\ref{sec:results} we present the main results of the paper, comparing predictions from the simulations with the properties of high-redshift star cluster candidates.
Finally, Section~\ref{sec:discussion} discusses limitations and biases that may affect the comparisons and Section~\ref{sec:summary} summarises the paper.

\section{Methods}
\label{sec:methods}

\subsection{Simulations}
\label{sec:simulations}

E-MOSAICS is a suite of cosmological hydrodynamical simulations of galaxy formation which include subgrid models for the formation and evolution of star clusters \citep{P18, K19a}, with the overall aims of testing the formation of GC populations and their use as tracers of the galaxy formation and assembly process.
The simulations couple the MOSAICS model for star cluster formation and evolution \citep{Kruijssen_et_al_11, P18} to the EAGLE (Evolution and Assembly of GaLaxies and their Environments) galaxy formation model \citep{S15, C15}.
The E-MOSAICS suite includes both zoom-in simulations of Milky Way-mass galaxies \citep{P18, K19a} and periodic cosmological volumes \citep{Pfeffer_et_al_19b, Bastian_et_al_20}.

The simulations were performed with a highly-modified version of $N$-body, smooth particle hydrodynamics code \gadget\ \citep{Springel_05}.
The EAGLE model includes subgrid routines for radiative cooling \citep[including the effect of the cosmic microwave background,][]{Wiersma_Schaye_and_Smith_09}, star formation \citep[where the effect of metal cooling and dust shielding is implemented as a metallicity-dependent density threshold for star formation, following \citealt{Schaye_04}]{Schaye_and_Dalla_Vecchia_08}, stellar evolution \citep{Wiersma_et_al_09}, the seeding and growth of supermassive black holes \citep{Rosas-Guevara_et_al_15} and feedback from star formation \citep{Dalla_Vecchia_and_Schaye_12} and black hole growth \citep{Booth_and_Schaye_09}.
Stellar feedback is implemented such that feedback is more efficient at higher gas densities and lower metallicities.
The feedback efficiencies were calibrated to reproduce the galaxy stellar mass function, galaxy sizes and black hole masses at $z \approx 0$ \citep{C15}.
The EAGLE simulations have been shown to broadly reproduce many properties of the evolving galaxy population, including the evolution of the galaxy stellar mass function and star formation rates \citep{Furlong_et_al_15}, galaxy sizes \citep{Furlong_et_al_17}, cold gas properties \citep{Lagos_et_al_15, Crain_et_al_17} and the galaxy mass--metallicity relation \citep{S15}.

As star clusters cannot be resolved in large cosmological simulations, the MOSAICS model treats star clusters as subgrid components of stellar particles.
Upon conversion of a gas particle to a star particle, the new star particle may form a subgrid population of star clusters based on local conditions in the simulation (gas properties and tidal field).
In MOSAICS, star cluster formation is controlled by two main functions: the cluster formation efficiency \citep[CFE or $\Gamma$, the fraction of stars formed in bound clusters,][]{Bastian_08} and shape of the initial cluster mass function (a power law, or \citealt{Schechter_76} function with an exponential upper mass truncation $\Mcstar$).
Initial cluster masses are drawn stochastically from the mass function, such that the subgrid clusters may be more massive than the stellar mass of the host particle.
In the fiducial model, the CFE traces the local natal gas pressure according to the \citet{Kruijssen_12} model (higher gas pressures result in more bound cluster formation), while the initial cluster mass function is a Schechter function where $\Mcstar$ varies according to the \citet{Reina-Campos_and_Kruijssen_17} model ($\Mcstar$ increases with gas pressure, except where limited by high Coriolis or centrifugal forces near the centres of galaxies).
Alternative models included in the simulations either fix the CFE to a constant value (10 per cent), assume no upper mass truncation (power law mass function), or both.
Following their formation, star clusters may lose mass at each timestep in the simulation from stellar evolution (according to the EAGLE model), two-body relaxation (depending on the local tidal field strength, \citealt{Lamers_et_al_05b, Kruijssen_et_al_11}; with an additional constant term to account for isolated clusters, \citealt{Gieles_and_Baumgardt_08}) and tidal shocks from rapidly changing tidal fields \citep{Gnedin_Hernquist_and_Ostriker_99, Prieto_and_Gnedin_08, Kruijssen_et_al_11}.
Additionally, dynamical friction is treated in post-processing at every snapshot and clusters are removed when the dynamical friction time-scale is less than the cluster age \citep[i.e. assuming they merge to the centre of their host galaxy, see][]{P18}.

In this work we analyse galaxies and their star clusters from the largest E-MOSAICS simulation, a periodic volume $34.4 \cMpc$ on a side which initially has $1034^3$ dark matter and gas particles \citep[L0034N1034,][]{Bastian_et_al_20}.
The dark matter and gas particles have initial masses of $m_\mathrm{dm} = 1.21 \times 10^6 \Msun$ and $m_\mathrm{b} = 2.26 \times 10^5 \Msun$, respectively, with Plummer-equivalent gravitational softening lengths of $1.33$~comoving~kpc prior to $z=2.8$, and are fixed to $0.35$~proper~kpc thereafter.
The simulation adopts cosmological parameters consistent with a \citet{Planck_2014_paperXVI} cosmogony ($\Omega_\mathrm{m} = 0.307$, $\Omega_\Lambda = 0.693$, $\Omega_\mathrm{b} = 0.04825$, $h = 0.6777$, $\sigma_8 = 0.8288$).
Snapshots were output for the simulation at 29 redshifts from $z=20$ to $z=0$.
Galaxies (subhaloes) were identified in the simulation snapshots in a two-step process. 
First, dark matter structures were detected with the friends-of-friends algorithm \citep{Davis_et_al_85}.
Next, bound subhaloes within each friends-of-friends group were identified using the \subfind\ algorithm \citep{Springel_et_al_01, Dolag_et_al_09}.
To connect descendant galaxies between snapshots, galaxy merger trees were constructed from the subhalo catalogues using the D-TREES algorithm \citep{Jiang_et_al_14, Qu_et_al_17}.

In this work we focus on the fiducial E-MOSAICS cluster formation model.
This model has been shown to be consistent with many scaling relations of present-day GC and young star clusters systems, such as 
the `blue tilt' of GC colour distributions \citep{Usher_et_al_18},
radial distributions of GC populations \citep{K19a, Reina-Campos_et_al_22a},
GC age-metallicity relations \citep{K19b, Kruijssen_et_al_20, Horta_et_al_21a},
scaling relations of young star clusters \citep{Pfeffer_et_al_19b},
the fraction of stars contained in GCs \citep{Bastian_et_al_20},
the high-mass truncation of GC mass functions \citep{Hughes_et_al_22}
and GC metallicity distributions \citep{Pfeffer_et_al_23b}.
However, the simulations overpredict the number of low-mass clusters \citep{P18}, as well as the number of high-metallicity GCs in Milky Way-mass galaxies \citep{Pfeffer_et_al_23b}.
This is potentially due to an overly-smooth interstellar medium \citep[EAGLE does not resolve the cold, dense phase of the interstellar medium,][]{S15}, resulting in insufficient disruption of star clusters by tidal shocks from dense gas clouds \citep[for detailed discussion, see][]{P18, K19a}.
This issue should not significantly affect the predominantly young ($\sim 10 \Myr$) and massive ($M \sim 10^6 \Msun$) clusters we are comparing in this work (see Section~\ref{sec:BrightestAgeMass}) as the timescales are too short for their disruption.

\subsection{Stellar population modelling}

To determine luminosities in \textit{JWST} NIRCam filters (focusing on F150W and F444W) for the simulated galaxies and star clusters in E-MOSAICS, we use the Flexible Stellar Population Synthesis (FSPS) model \citep{Conroy_Gunn_and_White_09, Conroy_and_Gunn_10}.
Stellar population luminosities were calculated from redshifted models using the redshift of simulation snapshots (i.e. we do not need to apply K-corrections as we are directly comparing observed-frame photometric bands).

We choose to compare observed-frame luminosities, rather than the more common approach of converting to a common rest-frame waveband, in order to remove dependencies of spectral energy distribution (SED) fitting on the observed luminosities.
In this way we can make as direct a comparison as possible that mainly depends on simple stellar population modelling for the simulations (with known ages and metallicities), rather than the degenerate (e.g. age-metallicity, star formation history), multi-parameter stellar population modelling required to fit observed galaxies.
However, comparisons of simulated and observed galaxies and star clusters are then only valid at the same redshift.

We assume the default FSPS parameters, and use the MILES spectral library \citep{Sanchez-Blazquez_et_al_06}, Padova isochrones \citep{Girardi_et_al_00, Marigo_and_Girardi_07, Marigo_et_al_08} and a \citet{Chabrier_03} initial stellar mass function (IMF, consistent with that used in the EAGLE model).
We calculate mass-to-light ratios for star particles and star clusters by linearly interpolating from the grid in ages and total metallicities.
As EAGLE does not model dust, we do not include dust extinction in the stellar population modelling, but instead correct the observed galaxies for extinction (Section~\ref{sec:observations}).

Each star particle and star cluster is assumed to be a simple stellar population formed in an instantaneous burst. 
To account for the formation timescale for star clusters of a few megayears \citep[e.g.][]{Chevance_et_al_20}, we also tested stellar populations formed with constant star formation rates over $5 \Myr$, but found our results (e.g. brightest cluster luminosities) consistent with using simple stellar populations.
Therefore, for simplicity, we adopt the predictions from simple stellar populations in the rest of this work.
We detail the luminosity selection of galaxies from the simulation in Section~\ref{sec:GalSelection}.

\subsection{Observations}
\label{sec:observations}

We compile from the literature a list of lensed galaxies observed with \textit{JWST} (predominantly with NIRCam) which contain compact star cluster candidates:

\begin{itemize}

\item The Cosmic Gems arc (SPT0615-JD) is a $z \approx 10.2$ galaxy lensed by a $z=0.972$ galaxy cluster \citep{Salmon_et_al_18, Adamo_et_al_24, Bradley_et_al_24}.
The arc contains five highly magnified ($\mu \approx 50$--$400$, with uncertainties $\approx 50$ per cent) compact sources with half-light radii $\approx 1 \pc$ or less, for which we adopt the properties from \citet{Adamo_et_al_24}.
For the lensed galaxy we adopt its properties from \citet{Bradley_et_al_24} and in particular adopt the photometric values of the counterimage, which was found to have a significantly higher intrinsic luminosity ($60$ per cent brighter) than the arc itself.
We also note that, based on \textit{HST} imaging, \citet{Salmon_et_al_18} found a higher stellar mass for the galaxy ($M_\ast = 2.0^{+2.0}_{-0.7} \times 10^{8} \Msun$) than that found by \citet[$M_\ast = 2.4$--$5.6 \times 10^7 \Msun$]{Bradley_et_al_24}.

\item The Firefly Sparkle arc is a $z=8.3$ galaxy lensed by a $z = 0.545$ galaxy cluster \citep{Postman_et_al_12, Schmidt_et_al_16, Hoag_et_al_17, Mowla_et_al_24}.
The arc contains ten compact sources with magnifications of $\mu \approx 16$--$26$ (with uncertainties $\approx 25$ per cent) and half-light radii $\lesssim 7 \pc$ \citep{Mowla_et_al_24}.
We adopt the properties for the arc and compact sources from \citet{Mowla_et_al_24}, although we note that \citet[using \textit{HST} and \textit{Spitzer} imaging]{Hoag_et_al_17} found a higher stellar mass for the galaxy ($M_\ast = 3.0^{+1.5}_{-0.8} \times 10^8 \Msun$) than that found by \citet[$M_\ast = 6.3^{+23.9}_{-2.8} \times 10^{6} \Msun$]{Mowla_et_al_24}.

\item The MACS J0416 arc is a $z=6.143$ system lensed by a $z=0.396$ galaxy cluster \citep{Caminha_et_al_17, Vanzella_et_al_17b, Vanzella_et_al_19}. The system contains three subsystems (D1, T1 and UT1) that may be three separate galaxies, each of which contains a compact source with half-light radius $<8 \pc$ \citep{Messa_et_al_24b}. Magnifications for the galaxies are in the range $\mu \approx 17$--$21$. We adopt the properties of the arcs and compact sources from \citet{Messa_et_al_24b}, treating each subsystem as a separate galaxy.

\item The Cosmic Grapes is a $z=6.072$ galaxy lensed by a $z=0.43$ galaxy cluster \citep{Fujimoto_et_al_21, Fujimoto_et_al_24, Laporte_et_al_21}. The galaxy is unique in that, though the magnification is high ($\mu \approx 32$, with uncertainties $\approx 3$ per cent), the distortion and differential magnification are minimal. The galaxy contains 15 sources with half-light radii of $\approx 7$--$60 \pc$. We adopt the properties of the galaxy and compact sources from \citet{Fujimoto_et_al_24}.

\item The Sunrise arc (WHL 0137–zD1) is a $z = 6 \pm 0.2$ galaxy lensed by a $z=0.566$ galaxy cluster, with magnifications of $\mu \sim 60$--$250$ along the arc \citep{Salmon_et_al_20, Welch_et_al_23}.
The arc contains six compact sources (three star cluster candidates and thee star-forming `nebular knots') with half-light radii $\approx 1$--$25 \pc$  \citep{Vanzella_et_al_23}.
We adopt the properties of the total arc and the three star cluster candidates from \citet{Vanzella_et_al_23}, noting that the reported magnifications are lower limits and thus the luminosities and masses of the star cluster candidates provide upper limits.

\item The Abell 2744 ``System 3'' arc is a $z=3.98$ galaxy lensed by a $z=0.308$ galaxy cluster \citep{Johnson_et_al_14, Mahler_et_al_18}.
The arc contains three compact sources with magnifications $\mu \approx 30$--$100$ and half-light radii $\approx 3$--$15 \pc$ \citep{Vanzella_et_al_22b, Bergamini_et_al_23}.
We adopt the properties of the arc and star cluster candidates from \citet{Vanzella_et_al_22b}, noting that this work used NIRISS imaging, rather than NIRCam imaging as for the other arcs.
For this galaxy we use the reported luminosities from the NIRISS F200W band, as the results for the F150W and F200W bands are very similar.

\item The Sparkler (SMACS0723 System 2) is a $z=1.378$ galaxy lensed by a $z=0.388$ galaxy cluster \citep{Golubchik_et_al_22, Caminha_et_al_22, Mahler_et_al_23}.
The galaxy contains 28 sources with half-light radii ranging from less than $10$ to a few $100 \pc$ \citep{Mowla_et_al_22, Claeyssens_et_al_23}.
For this galaxy, different lensing models have not yet converged on a solution for the galaxy cluster, with reported magnifications in the range $\mu \approx 10$--$100$ for image S2.2 and $\mu \approx 5$--$10$ for image S2.1 \citep{Caminha_et_al_22, Mahler_et_al_23, Chow_et_al_24}.
We adopt the properties of the star cluster candidates from \citet{Claeyssens_et_al_23}.
We note that \citeauthor{Claeyssens_et_al_23} assumed the \citet{Mahler_et_al_23} lensing model (giving $\mu \approx 10$ for S2.2), but if the \citet{Caminha_et_al_22} model were assumed, the magnification may be a factor $\sim 10$ higher, leading to smaller sizes, luminosities and masses for the star cluster candidates.
By adopting the lensing model with the lowest magnifications, these star cluster properties can be considered upper limits.
In addition to lens modelling uncertainty, there is significant uncertainty in the physical properties of many of the sources depending on the methods used for spectral energy distribution modelling \citep[see][]{Mowla_et_al_22, Adamo_et_al_23, Claeyssens_et_al_23}.
We adopt the physical properties of the galaxy from \citet{Mowla_et_al_22}.
For the luminosity of the galaxy we use photometry of the counterimage S2.1 from the \textit{JWST} Early Release Observations programme \citep{Pontoppidan_et_al_22} that is publicly available on the Dawn JWST Archive (DJA)\footnote{\url{https://dawn-cph.github.io/dja/imaging/v7/}}, as S2.2 (the most highly magnified image) is partially obscured by a foreground galaxy and \citet{Bradley_et_al_24} found the main arcs may underestimate the total luminosity of the galaxy.
Basic details of the data reduction for the DJA photometric catalogue are presented in \citep{Valentino_et_al_23}.
For consistency with the lensing model used by \citet{Claeyssens_et_al_23} we assume a total magnification $\mu = 5.1$ for S2.1 \citep{Mahler_et_al_23}.

\item SMACS0723 lensing cluster:
\citet{Claeyssens_et_al_23} presented measurements of compact sources in 18 lensed galaxies behind the lensing cluster SMACS0723 (which includes the Sparkler).
We include six of these systems (S1, S3, S4, S5, S7, I8, at redshifts $z = 1.449$, $1.991$, $2.19$, $1.425$, $5.173$, $2.12$ and with magnifications $\mu \approx 9.8$, $6.9$, $13.9$, $19.0$, $26.4$, $10.0$, respectively) that have compact sources satisfying our star cluster selection (see Section~\ref{sec:ClusterSelection}).
In cases where there are multiple images, we only include cluster candidates from the highest magnification image of each lensed galaxy, which have the highest number of compact sources (S1.2, S3.3, S4.2, S5.1, S7.1).
For three of the systems (S1, S4, S5) we use photometric measurements of the galaxies from the DJA (see above).
As for the Sparkler, we adopt the fluxes of the galaxies from the lowest magnified image available in the catalogue in each case (S1.3, S4.1, S5.3).
We adopt magnification estimates from \citet{Mahler_et_al_23} for consistency with \citet{Claeyssens_et_al_23}.
The magnifications have typical uncertainties of $\approx 20$ per cent.
The other three galaxies were not found in the catalogue, but we include them in the brightest cluster analysis for reference (Section~\ref{sec:Brightest}).
See \citet{Forbes_and_Romanowsky_23} and \citet{Adamo_et_al_23} for comparisons of Sparkler GCs and those of the Milky Way.

\end{itemize}

In addition to the above galaxies, when comparing the ages and masses of star cluster candidates (Section~\ref{sec:BrightestAgeMass}) we also include lensed galaxies which have been analysed with multiband \textit{HST} imaging, although we note that their properties are likely more uncertain due to the smaller wavelength range for SED fitting:

\begin{itemize}

\item The Sunburst arc is a $z=2.37$ galaxy lensed by a $z=0.443$ galaxy cluster \citep{Dahle_et_al_16} with a stellar mass of $\approx 10^9 \Msun$ \citep{Vanzella_et_al_22a}. The arc contains at least 12 compact sources with magnifications of $\mu \approx 15$--$500$ (with uncertainties $\approx 15$ per cent; though \citealt{Sharon_et_al_22} often find significantly lower magnifications by factors of up to $\approx 4$) and half-light radii of $3$--$20 \pc$, for which we adopt the properties from \citet{Vanzella_et_al_22a}. 

\item \citet{Messa_et_al_24} investigated the stellar clumps in three lensed galaxies using \textit{HST} imaging. Two of these, the RCS0224 \citep[$z=4.88$,][]{Gladders_et_al_02} and MACS0940 \citep[$z=4.03$,][]{Leethochawalit_et_al_16} arcs, contain compact clumps consistent with being star clusters (see Section~\ref{sec:ClusterSelection}).

\end{itemize}

For lensed galaxies and star cluster candidates we adopt the magnifications from the relevant works above.
However, we note that there can be systematic offsets in magnification depending on different lensing models that are much larger than the uncertainties from a particular model (e.g. up to a factor $\approx 10$ as discussed above for the Sparkler; or a factor $\approx 2$ between models for the Cosmic Gems tested by \citealt{Adamo_et_al_24}, larger than the $\approx 20$--$60$ per cent uncertainties from their reference model).

We correct the luminosities of the clumps and galaxies for internal dust extinction with their listed extinction values from SED fitting (where possible) using the \citet{Calzetti_et_al_00} attenuation relation \citep[following][]{Claeyssens_et_al_23}.
For the SMACS0723 lensed galaxies without extinction values, we use the median value of their clumps from \citet{Claeyssens_et_al_23}.
We summarise the properties of all galaxies in Table~\ref{tab:observations}.

\subsection{Galaxy selection}
\label{sec:GalSelection}

\begin{figure*}
  \includegraphics[width=0.495\textwidth]{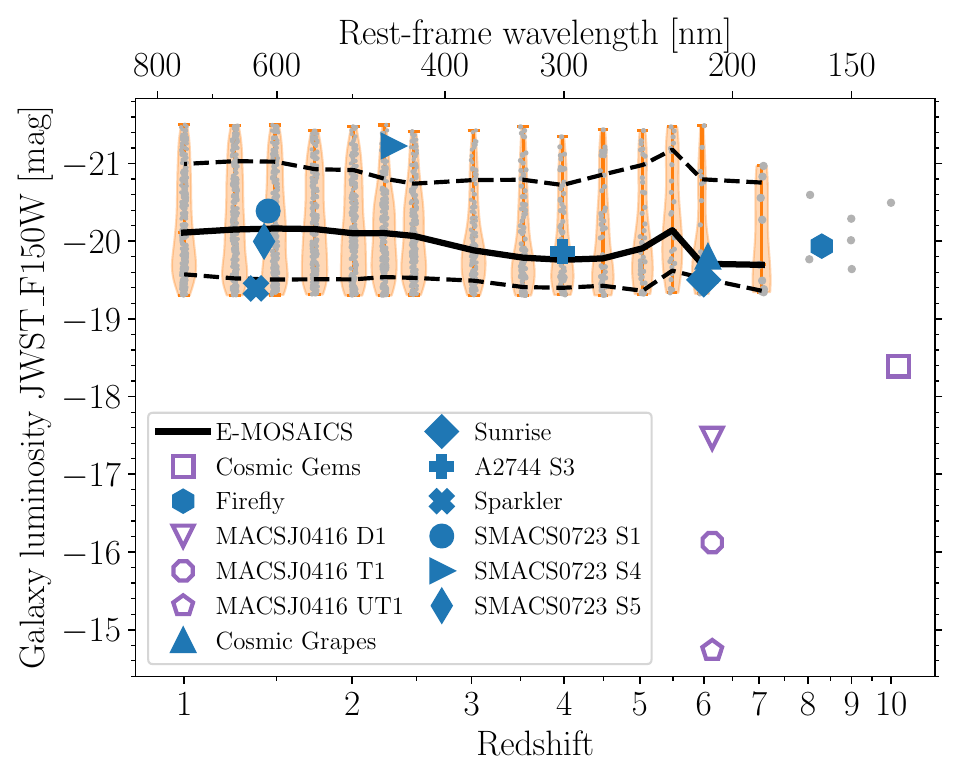}
  \includegraphics[width=0.495\textwidth]{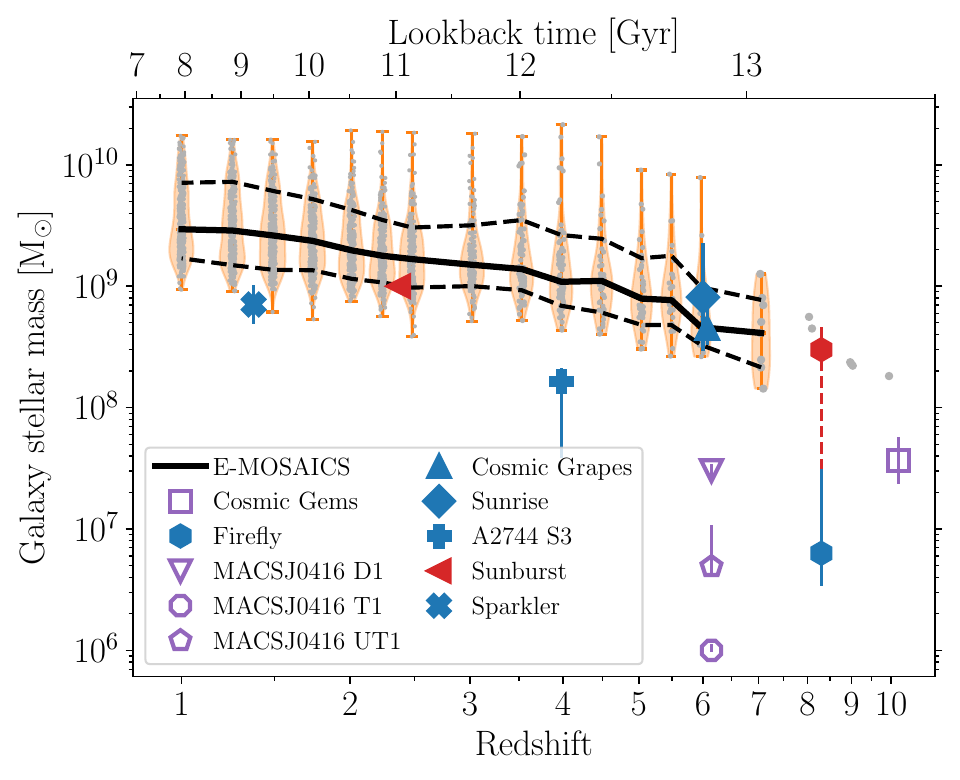}
  \caption{Comparison of the F150W absolute magnitudes (left) and stellar masses (right) of galaxies at different redshifts given the F150W luminosity selection to match the majority of observed lensed galaxies with compact sources ($-19.3 > M_\mathrm{gal,F150W} > -20.6$). E-MOSAICS galaxies at each snapshot redshift are shown as grey points (though with significant overlap at lower redshifts), with violin plots showing the distribution of values for each snapshot with at least 5 galaxies. Black solid and dashed lines show the median and 68 percentile range for the E-MOSAICS galaxies, respectively. Blue and purple markers show luminosities and mass estimates for lensed high-redshift galaxies with \textit{JWST} photometry (Cosmic Gems: \citealt{Adamo_et_al_24}; Firefly Sparkle: \citealt{Mowla_et_al_24}; Sunrise \citealt{Vanzella_et_al_23}; Cosmic Grapes: \citealt{Fujimoto_et_al_24}; A2744 System 3: \citealt{Vanzella_et_al_22b}; Sparkler: \citealt{Mowla_et_al_22}; MACS J0416 D1, T1 and UT1: \citealt{Messa_et_al_24b}). The Sunburst \citep{Vanzella_et_al_22a} galaxy (red triangle) was analysed with \textit{HST} photometry, but is shown in the mass panel as it also matches the simulated galaxy masses. Observed galaxies within the luminosity selection (as well as the Sunburst galaxy) are shown by filled markers and all others are shown by empty markers. For the Firefly galaxy we also show the mass estimate from \citet[using \textit{HST} and \textit{Spitzer} imaging]{Hoag_et_al_17} as a red hexagon. For reference, the top axis in the left panel shows the rest-frame wavelength of the F150W band at each redshift, while the top axis of the right panel shows the lookback time.}
  \label{fig:GalMass}
\end{figure*}

We focus on the NIRCam F150W band as it is common between most sources and available across the whole redshift range $1$--$10$, since the highest redshift sources are not detected in bluer bands due to the Lyman-$\alpha$ break \citep{Adamo_et_al_24}.
We apply a selection for the simulated galaxies in observed-frame F150W that largely encompasses the lensed galaxies with \textit{JWST} photometry, with the exception of the Cosmic Gems and MACSJ0416 D1, T1 and UT1 galaxies (which are much fainter than other galaxies), shown in the left panel of Fig.~\ref{fig:GalMass}.
Limiting the simulated galaxies to those resolved with stellar masses $M_\ast > 10^{7.5} \Msun$ ($>100$ star particles) corresponds to a luminosity limit of $M_\mathrm{F150W} \lesssim -18.5 \Mag$.
This means the MACSJ0416 galaxies are too faint for direct comparisons.
In principle the simulations can marginally resolve galaxies similar to the Cosmic Gems, but the simulation volume is too small for a large enough sample of galaxies at $z=10$.
We therefore apply luminosity limits of $-19.3 > M_\mathrm{gal,F150W} > -21.5$ at all redshifts in order to capture the luminosity range of the brighter lensed galaxies.
At redshifts $z \leq 3$ there are more than $100$ simulated galaxies at each snapshot, reaching $327$ galaxies by $z = 1$.
At redshifts $z \geq 5$ there are fewer than $50$ simulated galaxies at each snapshot, and fewer than $20$ galaxies at $z \geq 7$ (hence violin plots in Figure~\ref{fig:GalMass} are only shown for $z \leq 7$).
At these high redshifts ($z \gtrsim 5$) the luminosity function is not well sampled due to the limited simulation volume, leading to large changes in the median galaxy luminosity from snapshot to snapshot.

The right panel of Fig.~\ref{fig:GalMass} compares the stellar masses of simulated and observed galaxies (where mass estimates are available) in the luminosity range as a function of redshift.
This selection means all simulated galaxies are well resolved with $\gtrsim 1000$ stellar particles.
The larger stellar masses of the selected galaxy sample towards lower redshifts are a result of higher mass-to-light ratios, due to older stellar populations in galaxies at lower redshifts.
Given the uncertainties in stellar population modelling for luminosity selection (simulations) and deriving galaxy masses from SED fitting (observations), the simulated and observed galaxy mass ranges agree reasonably well for those with similar luminosities.
The observed galaxy mass estimates tend to be slightly lower than the simulated galaxy masses, which could indicate a tendency for younger ages in SED fitting or underestimated luminosities for the simulated galaxies.
However, the three brightest galaxies at $z<2.5$ (SMACS0723 S1, S4 and S5) currently do not have stellar mass estimates.
Based on the simulations we would expect stellar masses of $\sim 10^{9.5}$--$10^{10} \Msun$ for these galaxies.
The much lower mass of the Firefly Sparkle relative to other galaxies of similar luminosity is most likely due to the fitting of a top-heavy stellar IMF \citep{Mowla_et_al_24} relative to the \citet{Chabrier_03} IMF used in the EAGLE model, given it is one of the brightest of the lensed galaxies at $z>3$ (left panel of Fig.~\ref{fig:GalMass}).
Using a \citeauthor{Chabrier_03} IMF, \citet{Hoag_et_al_17} found a stellar mass for the galaxy of $M_\ast \approx 3 \times 10^8 \Msun$ (shown as the red hexagon in the figure), which would agree well with the simulated galaxy masses given its luminosity.
The four galaxies fainter than the luminosity selection (Cosmic Gems and MACSJ0416 D1, T1 and UT1) all have lower masses than the simulated galaxies, but generally agree with the mass-luminosity trends of the simulated galaxies at each redshift.

\begin{figure}
  \includegraphics[width=\columnwidth]{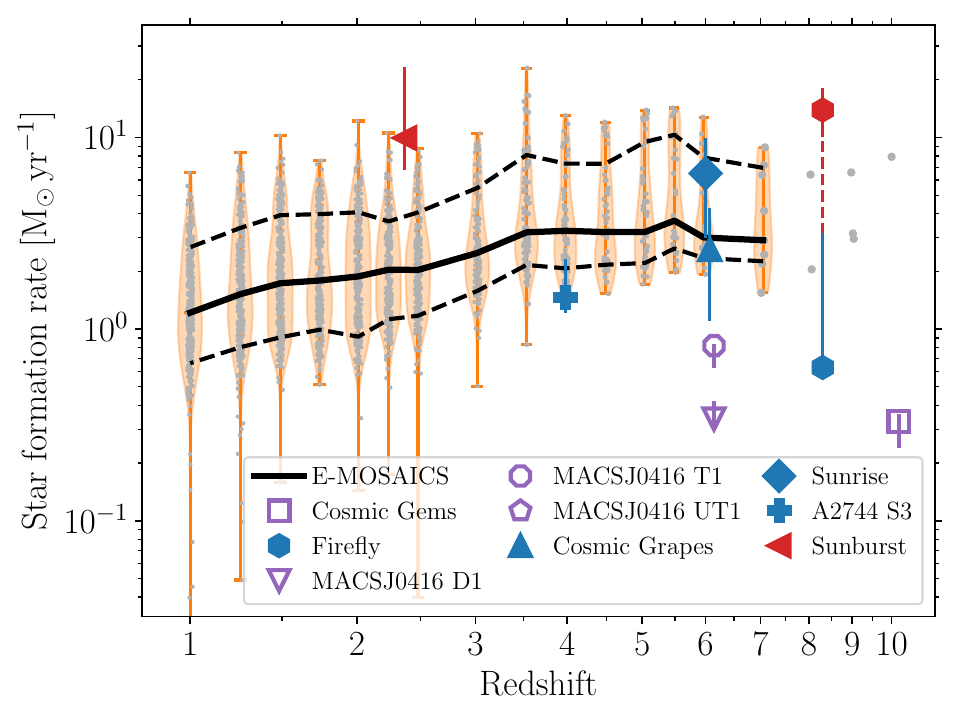}
  \caption{Star formation rates for the simulated and observed galaxies in Fig.~\ref{fig:GalMass}. Point and line styles are as in Fig.~\ref{fig:GalMass}. For Firefly we show SFR estimates from both \citet[blue point]{Mowla_et_al_24} and \citet[red point]{Hoag_et_al_17}.}
  \label{fig:GalSFR}
\end{figure}

Fig.~\ref{fig:GalSFR} shows the star formation rates (SFRs) for the same galaxies in Fig.~\ref{fig:GalMass} (where available for the observed galaxies).
\citet{Furlong_et_al_15} previously investigated SFRs in the EAGLE model as a function of redshifts, finding that, though the evolution as a function of redshift agrees well, EAGLE galaxies tend to underpredict SFRs (depending on the observational dataset).
For the available sample of lensed galaxies in Fig.~\ref{fig:GalSFR}, we find that the simulated and observed SFRs are in reasonable agreement.
Whether this is due to the small sample size, differences in the methods for determining SFRs of observed galaxies or the luminosity selection of galaxies in this work is not clear and would require a far larger sample size to test.

The selected simulated galaxies are generally the progenitors of relatively massive galaxies at $z=0$.
Following the selected galaxies through their merger trees to their descendants at $z=0$, the $z \geq 5$ galaxies are progenitors of $M_\ast(z=0) \approx 10^{11} \Msun$ galaxies, $z = 3$ galaxies are progenitors of $M_\ast(z=0) \approx 10^{10.5} \Msun$ galaxies and $z = 1$ galaxies are progenitors of $M_\ast(z=0) \approx 10^{10} \Msun$ galaxies (with typical $1 \sigma$ scatter in descendant stellar masses of $0.5$~dex).
Similarly, the $z>3$ galaxies are typically found in haloes with masses $M_{200}(z=0) > 10^{13} \Msun$ at $z=0$, $z=3$ galaxies in $M_{200}(z=0) \approx 10^{12.7} \Msun$ haloes and $z=1$ galaxies in $M_{200}(z=0) \approx 10^{12.2} \Msun$ haloes (the lower $1 \sigma$ scatter is typically $0.6$~dex, while the upper range is set entirely by the most massive group in the volume with $M_{200}(z=0) \approx 10^{13.7} \Msun$).
However, due to the limited simulation volume and the lack of rarer environments that will become galaxy clusters with $M_{200} > 10^{13.7} \Msun$ at $z=0$ (see discussion of this point in Section~\ref{sec:discussion}), the simulation most likely underestimates the descendant masses of the highest redshift ($z \gtrsim 4$) galaxies.

A number of works have discussed the connection of high redshift lensed galaxies to present day galaxies \citep[e.g.][]{Adamo_et_al_23, Forbes_and_Romanowsky_23, Fujimoto_et_al_24}.
Based on the Milky Way stellar mass history derived in \citet{K19b}, the galaxy selection in Fig.~\ref{fig:GalMass} could reasonably encompass Milky Way-type progenitor galaxies in the redshift range $\approx 1$--$6$ (in particular, the upper end of the mass range at $z \approx 1$ and the lower end of the mass range at $z \approx 6$).
This is of course dependent on the particular galaxy formation model, and it must be kept in mind that the EAGLE model slightly undershoots the `knee' of the galaxy stellar mass function \citep[i.e. there are slightly too few galaxies with stellar masses $\approx 10^{10.5} \Msun$,][]{S15}.

\subsection{Star cluster selection}
\label{sec:ClusterSelection}

For star cluster properties we also focus on the NIRCam F150W band (rest-frame wavelength of $\approx 750~\mathrm{nm}$ at $z = 1$ to $\approx 140~\mathrm{nm}$ at $z = 10$) but additionally compare the simulations and observations in the redder band NIRCam F444W ($\approx 220~\mathrm{nm}$ at $z = 1$ to $\approx 400~\mathrm{nm}$ at $z=10$).
We select compact sources with magnification-corrected half-light radii (or upper limits for unresolved objects) $R_\mathrm{eff} < 20 \pc$, i.e. sizes consistent with star clusters, as \citet{Messa_et_al_19} found that clumps with sizes $> 20 \pc$ may contain multiple star clusters.
This size limit would capture essentially all young clusters in nearby galaxies \citep{Brown_and_Gnedin_21}.
The exception to this criterion is for the Sparkler where we also include sources in the GC candidate list from \citet{Adamo_et_al_23}, which have half-light radii up to $\approx 50 \pc$.
These sources are offset from the galaxy itself, where confusion with cluster complexes/star forming regions is not likely to be an issue.
We note that excluding the more extended sources does not change the brightest cluster comparisons (Section~\ref{sec:Brightest}) as the brightest cluster has $R_\mathrm{eff} < 12 \pc$.

For the E-MOSAICS galaxies, we include all star clusters in particles bound to the galaxies according to \subfind.
We exclude a small fraction of particles with very low metallicities ($\FeH < -3$) as they may strongly depend on the treatment of Population III stars (which are not modelled in EAGLE).

\section{Results}
\label{sec:results}

\subsection{Cluster luminosity fractions}
\label{sec:LumFrac}

\begin{figure*}
    \includegraphics[width=0.495\textwidth]{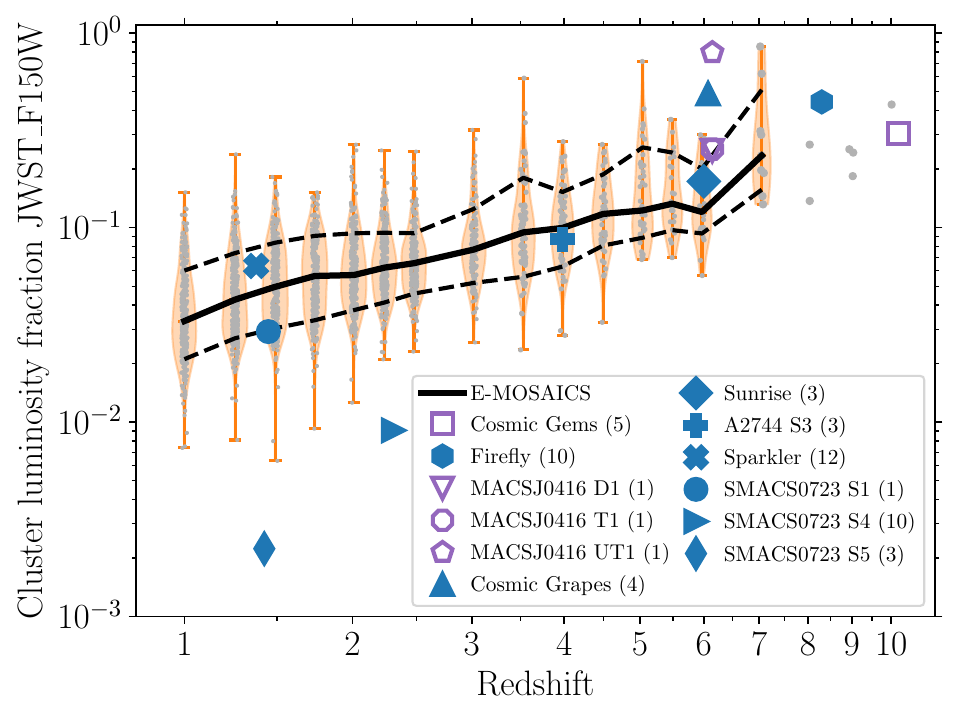}
    \includegraphics[width=0.495\textwidth]{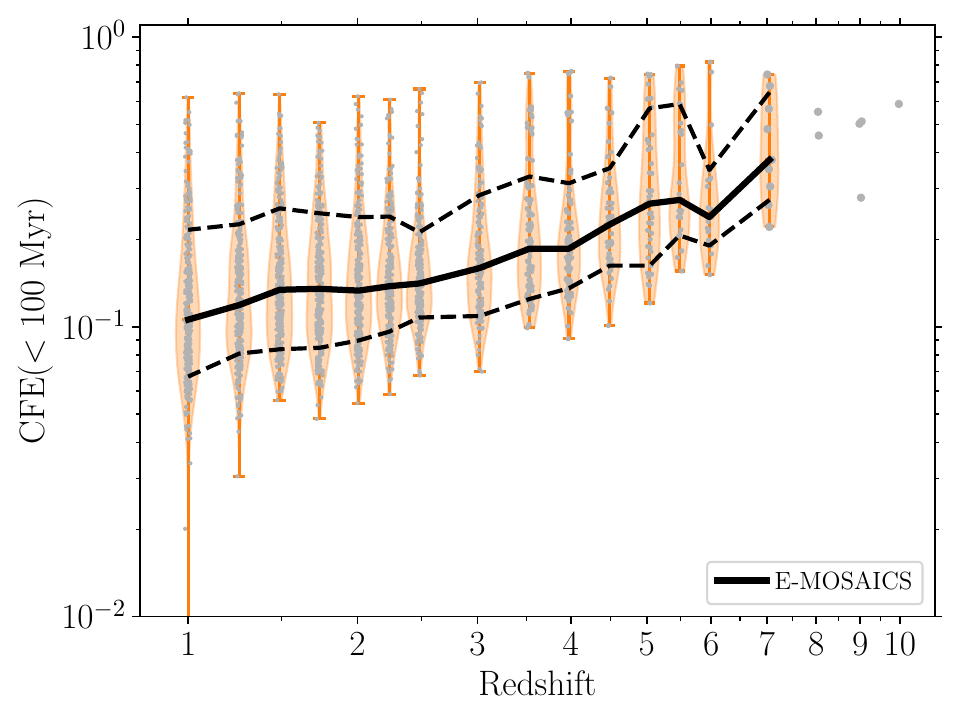}
    \caption{Total cluster luminosity fraction (i.e. all surviving star clusters in E-MOSAICS galaxies) in NIRCam F150W band (left) and cluster formation efficiency (CFE) in the last $100 \Myr$ (right) as a function of redshift for galaxies matching the luminosity selection. Point and line styles are as in Fig.~\ref{fig:GalMass}. The number of compact sources ($R_\mathrm{eff} < 20 \pc$) in each of the observed galaxies are noted in brackets in the caption.}
    \label{fig:LumFrac}
\end{figure*}

As a first comparison, the left panel of Figure~\ref{fig:LumFrac} shows the luminosity fraction of high redshift galaxies that is contributed by star clusters in both simulated E-MOSAICS galaxies and observed high-redshift lensed galaxies.
For the simulations we include all surviving clusters in the luminosity fractions (typically a few thousand clusters), but the brightest five clusters generally contribute 10--30 per cent of the total cluster light increasing to $\approx 80$ per cent in some cases.
In fact, we find the fraction of cluster light in the brightest few clusters decreases with decreasing redshift due to the larger contribution of older, fainter clusters in older galaxies.
The brightest 5 (10) clusters in F150W typically contribute $\approx 30$ (40) per cent of cluster light at $z=7$ and $\approx 10$ (13) per cent at $z=1$.
The fraction is also slightly larger in bluer filters due to faster fading of stellar populations.

We note that only a handful of compact sources are generally detected in the observed lensed galaxies (see caption in figure), compared with the steeply increasing cluster luminosity functions found in low redshift galaxies \citep[e.g.][]{Whitmore_and_Schweizer_95, Larsen_02}, implying only the brightest few clusters are detected and the luminosity fractions are lower limits.
However, the luminosity fractions are sensitive to resolution, meaning, in cases where star clusters are unresolved, the clumps may be blended star-forming regions or multiple clusters, leading to overestimated luminosity fractions \citep{Cava_et_al_18, Messa_et_al_19}.
In the opposite sense, clumps excluded for having sizes much larger than star clusters ($R_\mathrm{eff} > 20 \pc$) likely still contain star clusters within them, though the fraction of light being contributed by star clusters to the clumps is unknown.
As such, this figure should only be taken as a qualitative comparison between the simulations and observations.
Direct comparisons require `re-observing' resolved galaxies (observed or simulated) with the lensing model and point spread function from each high-redshift galaxy \citep[c.f.][]{Messa_et_al_19, Vanzella_et_al_19}.

Overall, the simulations predict a decreasing trend of cluster luminosity fraction with decreasing redshift, from $\approx 20$ per cent at $z=7$ to $\approx 3$ per cent at $z=1$.
There is relatively large scatter from galaxy to galaxy, including some galaxies which approach a luminosity fraction of unity.
These later cases are due to very bright, young ($<10 \Myr$) clusters \citep{Pfeffer_et_al_19a}, meaning the luminosity fractions will decrease as the clusters fade.

Though the total cluster luminosity fractions from the simulations do not provide a direct comparison with the observed cluster luminosity fractions, the observed galaxies are consistent with a similar decrease in luminosity fractions with decreasing redshift.
We note that the luminosity fractions presented here for the lensed galaxies differ from those presented in previous works (Section~\ref{sec:observations}) due to the adoption of an upper size limit for clusters, extinction corrections for luminosities and the use of counterimages for total luminosities (where possible).
The lensed galaxies SMACS0723 S4 and S5 have cluster luminosity fractions well below the simulations and other observed galaxies, implying they may be particularly affected by the missing contribution of undetected clusters, even though both galaxies have the faintest detected clusters compared to other galaxies at similar redshifts (see Fig.~\ref{fig:Brightest}).
As we will see in the next section, both galaxies are in much better agreement with the simulations when only considering the brightest cluster.
The fainter galaxies outside our luminosity selection (Cosmic Gems, D1, T1, UT1) have similar cluster luminosity fractions to the more luminous galaxies, implying there may not be any strong trends with galaxy luminosity/mass.

Given the range of redshifts and rest-frame wavelengths, the causes of this decrease in the simulations are multifold, and include decreasing CFE with decreasing redshift and disruption of star clusters with time \citep[see also][]{P18, Pfeffer_et_al_19a, Bastian_et_al_20}.
The decreasing CFE is demonstrated in the right panel of Fig.~\ref{fig:LumFrac}.
The CFE decreases from $\approx 50$ per cent at $z>7$ (though the sample size is small) to $\approx 10$ per cent at $z=1$.
Similar mass ($10^9$--$2\times 10^{10} \Msun$) star-forming galaxies at $z=0$ have typical CFEs of $\approx 8^{+7}_{-3}$ per cent.
In general, the CFE is expected to be larger than the cluster luminosity fraction \citep{Pfeffer_et_al_19a}.
As galaxies evolve, the cluster luminosity fraction then decreases due to star cluster disruption and fading of older clusters.

\subsection{Brightest star clusters}
\label{sec:Brightest}

\begin{figure*}
  \includegraphics[width=0.495\textwidth]{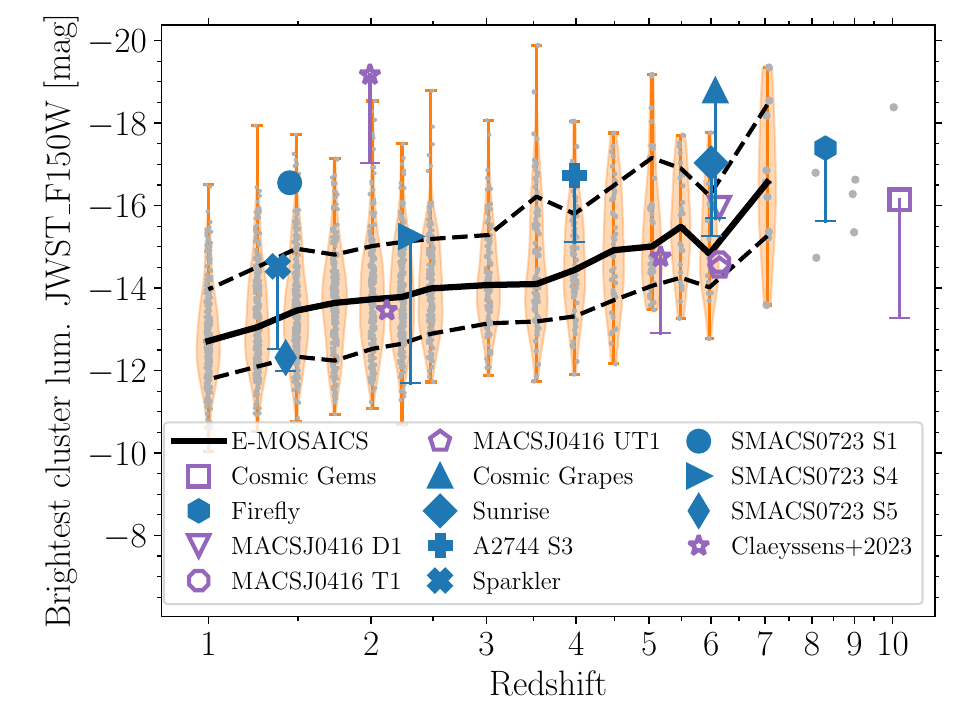}
  \includegraphics[width=0.495\textwidth]{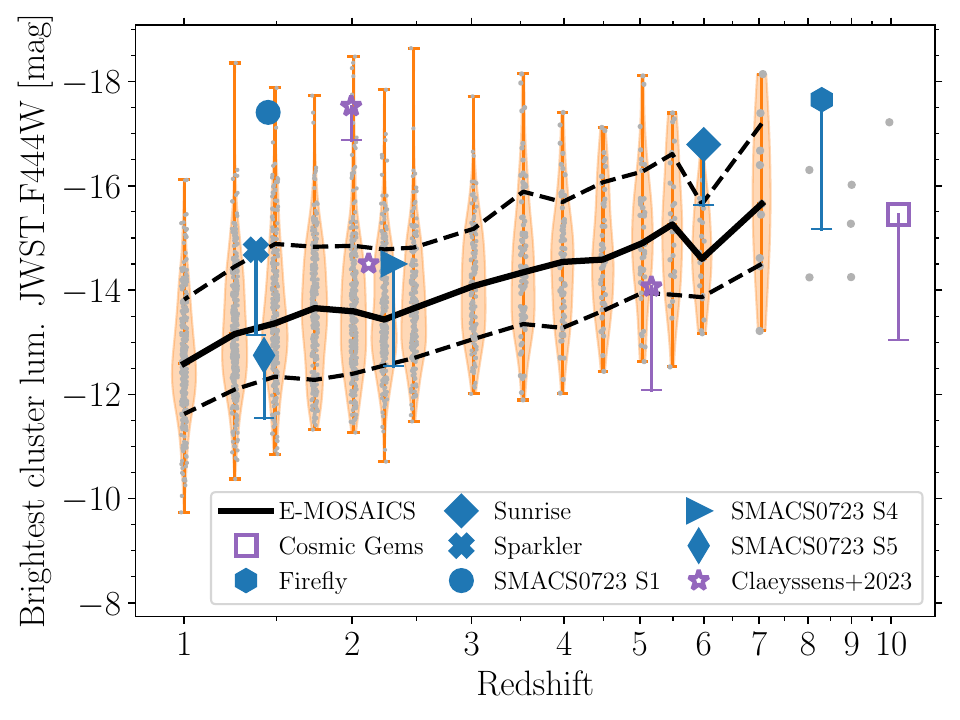}
  \caption{Luminosity of the brightest cluster in observed-frame NIRCam bands F150W (left) and F444W (right) as a function of redshift. Due to the changing rest-frame wavelengths, comparisons are only valid at a given redshift. Point and line styles are as in Fig.~\ref{fig:GalMass}. Purple stars show additional observations from \citet{Claeyssens_et_al_23} in galaxies without total luminosity estimates. The `lower limits' on the observed points are not errorbars, but instead show the faintest cluster candidate for each galaxy as a measure of detection limits in the observed samples.}
  \label{fig:Brightest}
\end{figure*}

As a more direct comparison, particularly for galaxies with very few compact sources, Fig.~\ref{fig:Brightest} shows the absolute luminosities of the brightest star clusters in each galaxy (redshift and magnification corrected for observed galaxies).
The left panel shows results for the brightest clusters in NIRCam F150W, while the right panel shows results for NIRCam F444W.
We stress here that, as these are observed-frame luminosities, the evolving rest-frame wavelengths with redshift mean that comparisons are only valid at a given redshift.
We find that the brightest cluster candidates for the observed galaxies are largely within the range of brightest clusters predicted for the E-MOSAICS galaxies at each redshift.
However, the brightest cluster candidates in observed lensed galaxies are generally brighter than the median for E-MOSAICS galaxies.
The offset appears independent of redshift, with the observed galaxies following closely the $68\ts{th}$ percentile of the simulated galaxies.
The results are similar for both F150W (left panel) and F444W (right panel), as well as other bands not shown (F090W, F200W).
We also tested the effect of not modelling dynamical friction for massive clusters but found the results were largely unchanged (with only some small changes in the bright-end scatter) due to their generally young ages (Fig.~\ref{fig:BrightestAge}) that are smaller than the dynamical friction timescales.

One possible reason for the brighter offset of observed galaxies could be detection limits of star clusters, even in strongly magnified galaxies.
To investigate this, in the figure we also indicate the luminosity of the faintest cluster candidate in each galaxy as `lower limits', though noting this will not be the same as the real detection limit.
Half ($4/8$ in F150W, $3/6$ in F444W) of the galaxies within the luminosity selection (Fig.~\ref{fig:GalMass}) have a faintest detected cluster that is similar to or brighter than the median luminosity for E-MOSAICS galaxies at similar redshifts, and nearly all have a faintest cluster that is brighter than `least luminous' brightest cluster in the simulated galaxies at comparable redshifts.
If the faintest clusters are similar to detection limits and the intrinsic brightest cluster distribution was similar to E-MOSAICS galaxies, this could imply around half of the galaxies would not have detectable clusters and would be excluded from the sample, leading to a bias towards galaxies with brighter clusters.
This would similarly apply to the luminosity fractions in Fig.~\ref{fig:LumFrac}.
We discuss this and other possible causes of differences in the observed galaxies and simulation predictions further in Section~\ref{sec:discussion}.

\subsection{Brightest cluster--SFR relation}
\label{sec:MV_brightest}

\begin{figure}
  \includegraphics[width=\columnwidth]{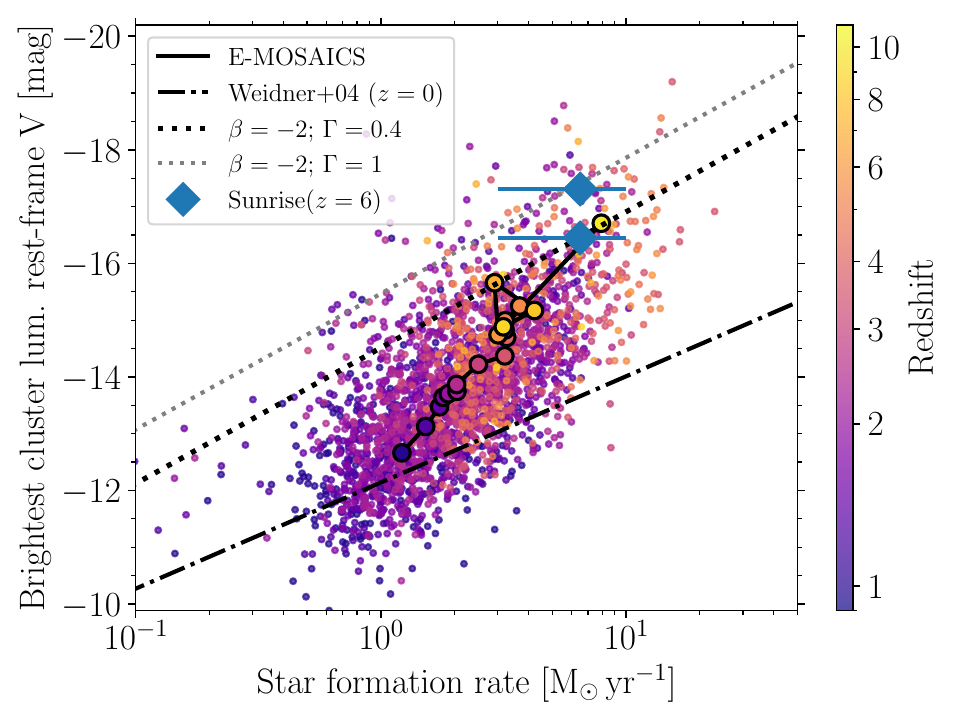}
  \caption{Relation between the brightest cluster in rest-frame $V$ band and galaxy SFR for high-redshift ($z\geq 1$) galaxies. Small coloured points show E-MOSAICS galaxies from the luminosity selection in Fig.~\ref{fig:GalMass}, coloured by snapshot redshift. The solid black line with large outlined circles shows the median SFR and brightest cluster for E-MOSAICS galaxies at each snapshot. The dash-dotted line shows the best-fitting relation for observed galaxies at $z \approx 0$ from \citet{Weidner_Kroupa_and_Larsen_04}. The black dotted line shows the expected relation for a power-law cluster mass function (index $\beta = -2$) with 40 per cent cluster formation efficiency \citep[$\Gamma = 0.4$,][]{Bastian_08}, while the grey dotted line shows the expected relation for $\Gamma = 1$. The blue triangles show the brightest cluster in \textit{JWST} bands F356W and F410W (both close to rest-frame $V$) from the Sunrise arc at $z \approx 6$ \citep{Vanzella_et_al_23}, which agrees well with the trend predicted by the simulations. Overall, the simulations predict steeper (than that found at $z=0$) brightest cluster--SFR relations at higher redshifts.}
  \label{fig:Brightest_SFR}
\end{figure}

In the low-redshift galaxy population, observations have found a correlation between the brightest cluster in the $V$ band and the star formation rate (SFR) of a galaxy \citep{Billett_Hunter_and_Elmegreen_02, Larsen_02, Weidner_Kroupa_and_Larsen_04}
The correlation is not simply a cluster size-of-sample effect (and thus dependent on the CFE), but is also sensitive to the upper truncation of the cluster mass function which affects the slope of the correlation \citep{Bastian_08}.
In \citet{Pfeffer_et_al_19b} we showed that the fiducial E-MOSAICS model agrees well with the observed relation at $z=0$ \citep{Weidner_Kroupa_and_Larsen_04}.

In Fig.~\ref{fig:Brightest_SFR} we test if such a correlation is still in place at higher redshifts, using the rest-frame $V$ band to factor out evolving rest-frame wavelengths.
The scatter in the brightest cluster at a given SFR for the simulated galaxies is due differences in CFE and upper cluster mass truncation ($\Mcstar$) between galaxies, as well as stochasticity in sampling the cluster mass functions and the SFRs (measured instantaneously from the SFRs of the gas particles), which is why some points lie above the implied upper limit of $\Gamma =1$ (grey dotted line).
From $z=1$ to $z\geq 5$, the median SFR in the simulated galaxies increases by a factor of three (see also Fig.~\ref{fig:GalSFR}).
However, the increase in the brightest cluster luminosity over the same period ($\approx 2.5 \Mag$) is far larger than that expected from the $z=0$ brightest cluster relation \citep[dash-dotted line,][]{Weidner_Kroupa_and_Larsen_04}.
This difference is due to the increase in the CFE, which increases by a factor of four from $z=1$ to $z=7$ due to increasing natal gas pressure (Fig.~\ref{fig:LumFrac}), and an increase in $\Mcstar$ with redshift due to higher gas fractions \citep{Pfeffer_et_al_19b}.
The highest redshift simulated galaxies ($z \geq 5$) approach the expected relation for a CFE of 40 per cent (the median CFE at $z=7$) with no upper mass truncation (black dotted line).
As redshift decreases, the simulated galaxy population converges to the observed $z=0$ relation, which is well fitted by a CFE of $\approx 8$ per cent \citep{Bastian_08}, similar to the CFE of $\approx 10$ per cent for the simulated galaxies at $z=1$ (Fig.~\ref{fig:LumFrac}).

In the figure we also compare high-redshift observations of the Sunrise arc at $z \approx 6$ \citep{Vanzella_et_al_23}, which has \textit{JWST} imaging (F356W and F410W) close to the rest-frame $V$ band (effective wavelength within the full-width-half-maximum of the $V$ filter).
The Sunrise arc agrees well with the trend predicted by E-MOSAICS galaxies, though it falls above the median for simulated $z=6$ galaxies ($\mathrm{SFR} \approx 3 \Msun \yr^{-1}$, $M_V \approx -15$).
Depending on the filter, a CFE of $\sim 40$--$100$ is expected for Sunrise based on its location in the brightest cluster--SFR diagram, which agrees with a CFE of $30$--$60$ per cent estimated by \citet{Vanzella_et_al_23} from its SFR surface density.
Further high-redshift observations could be compared by using SED fits to calculate rest-frame $V$ luminosities.

\subsection{Brightest cluster ages and masses}
\label{sec:BrightestAgeMass}

\begin{figure*}
  \includegraphics[width=0.495\textwidth]{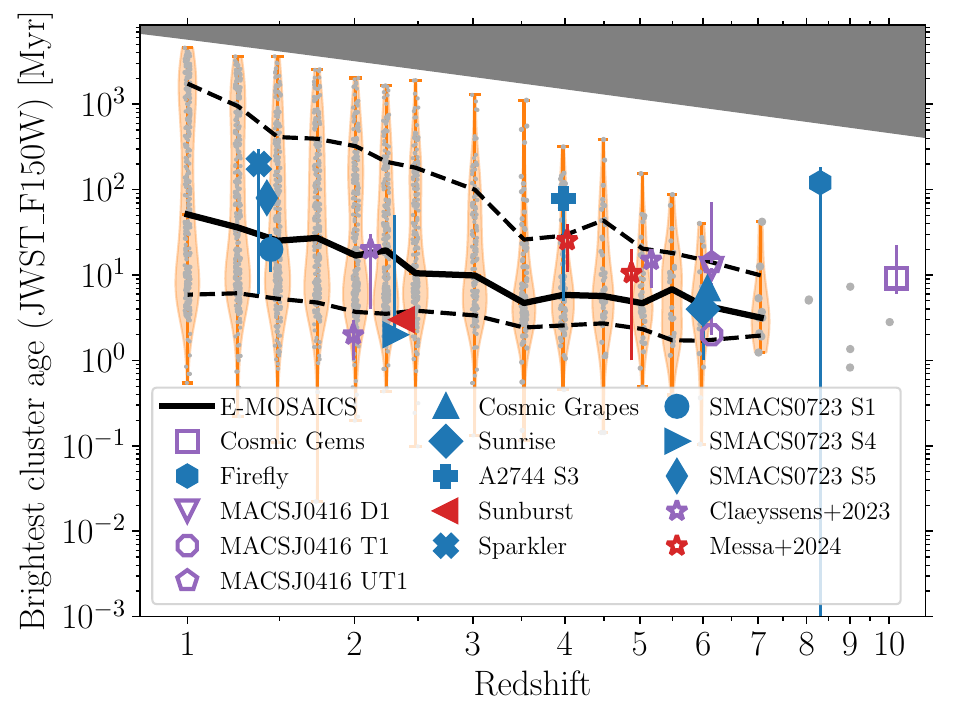}
  \includegraphics[width=0.495\textwidth]{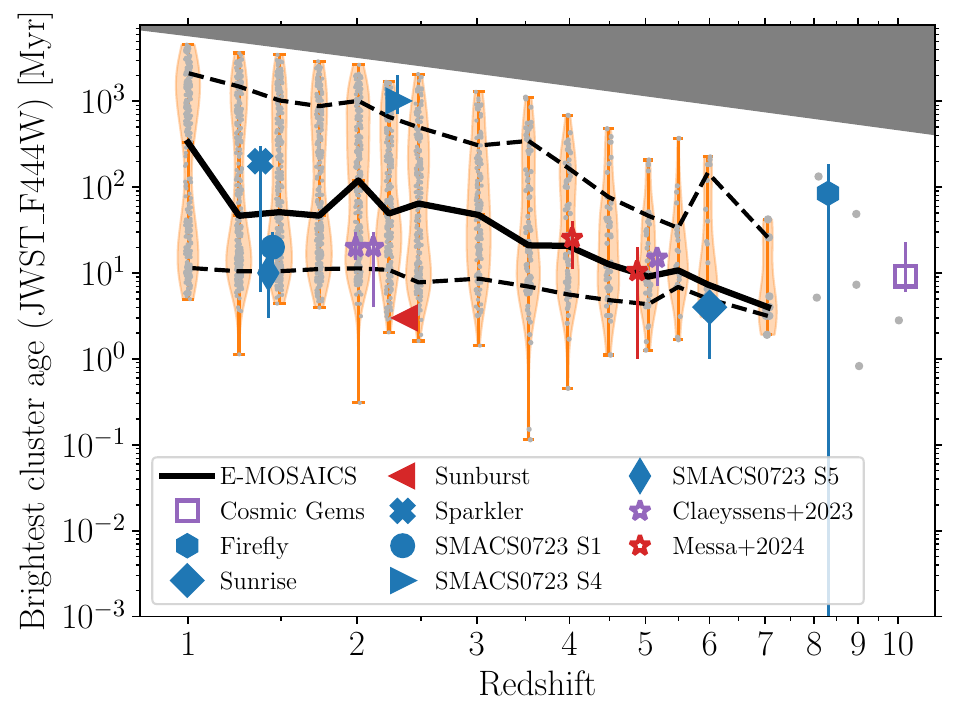}
  \caption{Ages of the brightest clusters in F150W (left) and F444W (right) as a function of redshift. Point and line styles are as in Fig.~\ref{fig:GalMass}. Here errorbars show the uncertainties in property estimates from SED fitting (see Section~\ref{sec:observations} for references). Purple stars show additional observations from \citet{Claeyssens_et_al_23} in galaxies without total luminosity estimates. The upper shaded regions show the limit from the age of the Universe. In the panels the red points include lensed galaxies with \textit{HST} imaging for comparison \citep{Vanzella_et_al_19, Vanzella_et_al_22a, Messa_et_al_24}.}
  \label{fig:BrightestAge}
\end{figure*}

In Fig.~\ref{fig:BrightestAge} we compare the ages of the brightest cluster in each simulated galaxy.
The oldest ages are of course limited by the age of the Universe, indicated by the grey shaded region in the left panel.
The brightest cluster ages are sensitive to stellar population fading, thus the brightest clusters are expected to be older in redder rest-frame bands \citep{Pfeffer_et_al_19a}.
They also depend on the maximum cluster mass and cluster formation rate over time \citep[size-of-sample effects,][]{Gieles_et_al_06b}, meaning they are also sensitive to galaxy and star cluster formation histories.

At $z>3$, we predict the median brightest clusters in F150W to have ages $\lesssim 10 \Myr$.
The median ages then increase towards lower redshifts, reaching $\approx 50 \Myr$ at $z=1$.
In F444W the trend is similar, but the median cluster ages are slightly older ($\approx 100 \Myr$ at $z=1$, as expected from the redder band).
The increasing ages towards lower redshifts are driven by the changing rest-frame wavelengths with redshift, along with the increasing age of the earliest formed clusters at lower redshifts.
At $z \lesssim 2.5$ the distribution of ages (violin plots) at each snapshot often appears bimodal between $\sim 10 \Myr$ and $\sim 10^3 \Myr$; i.e. young, very bright clusters or old, very massive clusters \citep[with the later most similar to some GC candidates in the Sparkler and SMACS0723 S4,][]{Mowla_et_al_22, Adamo_et_al_23, Claeyssens_et_al_23}.

The SED fitting-derived ages of cluster candidates from observed galaxies agree very well with the predictions from the simulated galaxies, generally falling within the $1 \sigma$ region (dashed lines).
Given the sensitivity of brightest cluster ages to the combination of a number of effects (stellar population evolution, cluster formation histories), the good agreement between the observed and simulated galaxies shows that similar star cluster formation processes (like the `young cluster' based model implemented in E-MOSAICS) may be operating in both low- and high-redshift galaxies.
We note, though, that ages from SED fitting can be sensitive to methodology.
For example, in Fig.~\ref{fig:BrightestAge} we use cluster ages for the Sparkler from \citet{Claeyssens_et_al_23}, as they provide the largest sample of clump properties. 
Many of the star cluster candidates were also analysed by \citet{Mowla_et_al_22} and \citet{Adamo_et_al_23}, who found significantly different ages in some cases.

\begin{figure}
  \includegraphics[width=\columnwidth]{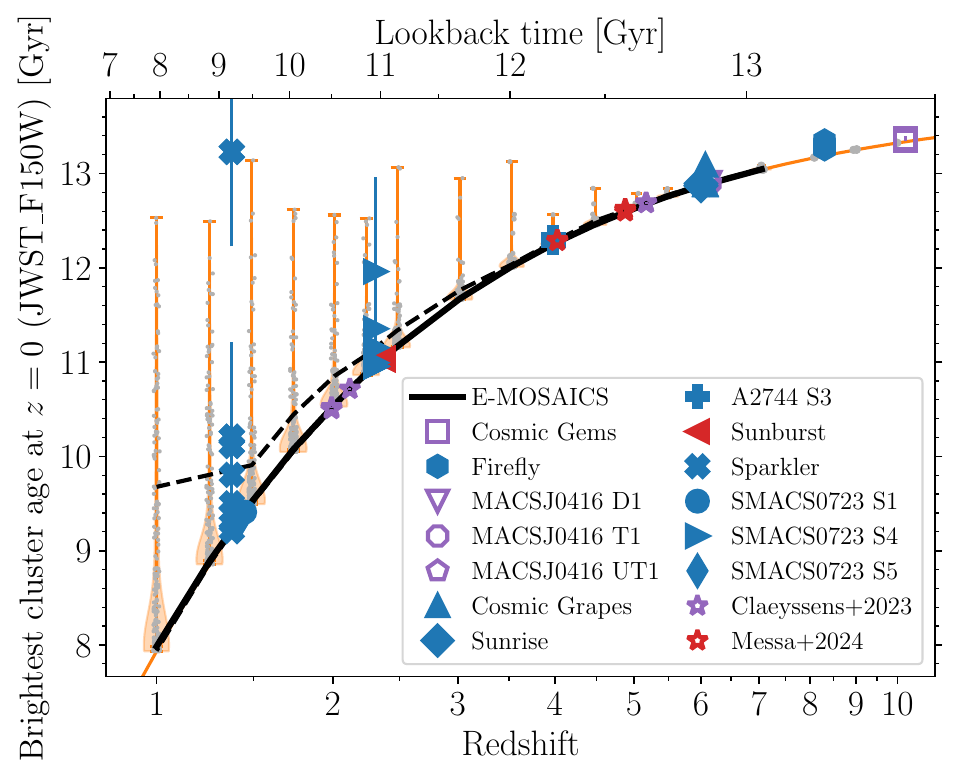}
  \caption{Ages of the simulated brightest clusters in F150W at $z=0$ (i.e. lookback time of formation) as a function of redshift. Point and line styles are as in Fig.~\ref{fig:GalMass}. The minimum possible age at each redshift (i.e. the lookback time at that redshift) is indicated by the orange curve, though is largely covered by the median E-MOSAICS prediction (solid black line). For the observed galaxies, we show the ages of \textit{all} cluster candidates, rather than only the brightest cluster.}
  \label{fig:BrightestFormTime}
\end{figure}

As an alternative way to view their ages, in Fig.~\ref{fig:BrightestFormTime} we compare the ages at $z=0$ for the brightest cluster in E-MOSAICS galaxies (i.e. assuming that the clusters survive until $z=0$ to become GCs).
Given that their `observed' ages at each redshift are $\lesssim 100 \Myr$ (Fig.~\ref{fig:BrightestAge}), for most clusters their $z=0$ ages are similar to the lookback time at each snapshot redshift.
For comparison we show the $z=0$ ages of \textit{all} star cluster candidate in the lensed galaxies in our sample.
Similar to the simulated galaxies, there are only a few star cluster candidates (namely, from the Sparkler and SMACS0723 S4) with ages much larger ($\gtrsim 1 \Gyr$) than the lookback time at which they are observed.
This can be explained by the fading of stellar populations as they age: even in rest-frame visible and near infrared bands, young clusters ($\lesssim 100 \Myr$) are generally expected to be the brightest and most readily-observable in high redshift galaxies \citep[see figure 2 of][]{Pfeffer_et_al_19a}.
Given the limit of the minimum possible $z=0$ age at each redshift, large age ranges ($\gtrsim 1 \Gyr$) in the populations only begin to occur at $z \lesssim 3$.
It also shows that `proto-GC' formation is occurring across a wide range of redshifts, rather than at a specific epoch \citep[like that required by models of GC formation in dark matter minihaloes, e.g.][]{Trenti_Padoan_and_Jimenez_15, Boylan-Kolchin_17, Creasey_et_al_19}.

\begin{figure*}
  \includegraphics[width=0.495\textwidth]{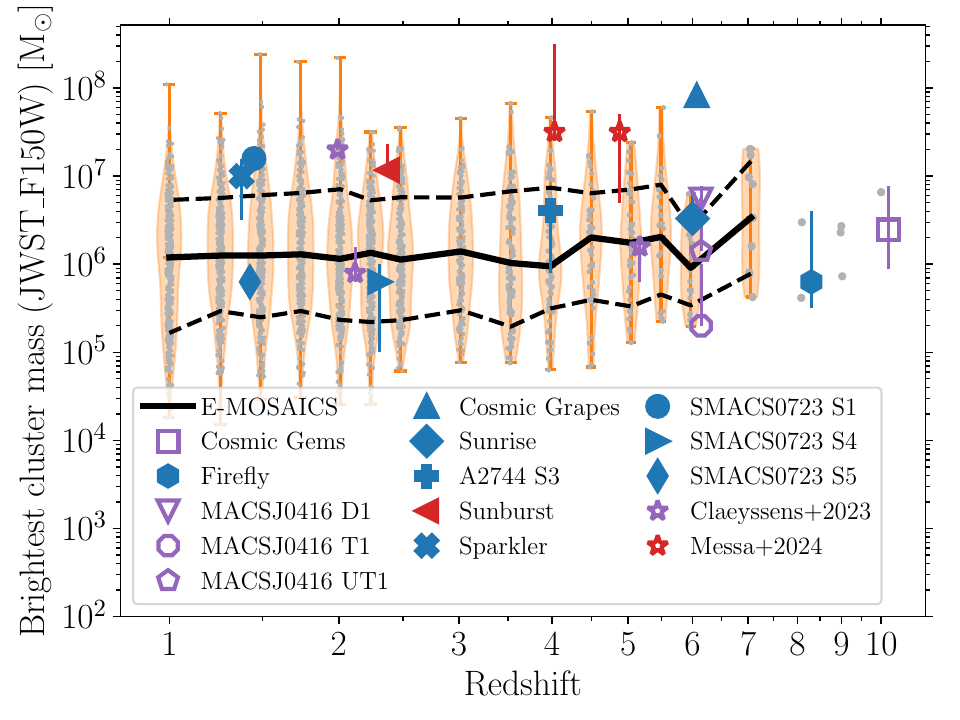}
  \includegraphics[width=0.495\textwidth]{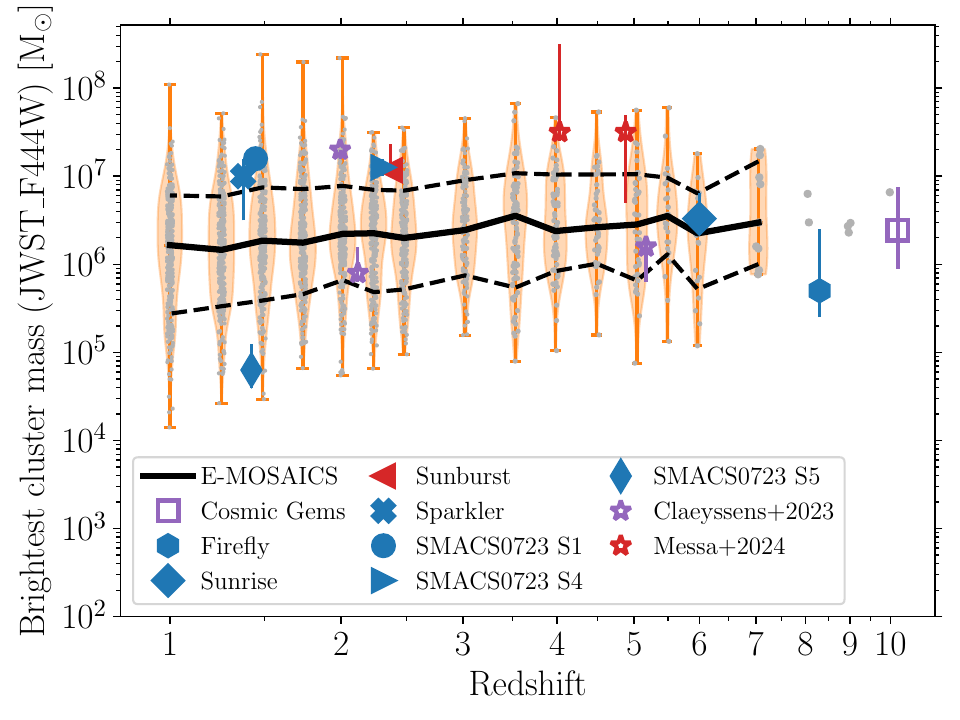}
  \caption{Stellar masses of the brightest clusters in F150W (left) and F444W (right) as a function of redshift. Point and line styles are as in Fig.~\ref{fig:GalMass}. Here errorbars show the uncertainties in property estimates from SED fitting (see Section~\ref{sec:observations} for references). Purple stars show additional observations from \citet{Claeyssens_et_al_23} in galaxies without total luminosity estimates. In the panels the red points include lensed galaxies with \textit{HST} imaging for comparison \citep{Vanzella_et_al_19, Vanzella_et_al_22a, Messa_et_al_24}.}
  \label{fig:BrightestMass}
\end{figure*}

Figure~\ref{fig:BrightestMass} shows the masses of the brightest clusters.
We predict a very flat distribution in median masses of $\approx 10^6 \Msun$, with the median masses in F444W marginally larger than those in F150W.
The bimodality found in cluster ages at $z \lesssim 2.5$ (Fig.~\ref{fig:BrightestAge}) is not found in the brightest cluster masses.
The mass of the brightest cluster is difficult to interpret in the context of the maximum mass of the cluster mass function, as it depends upon the cluster age distribution/formation history (size-of-sample effects) and rest-frame wavelength (rate of fading), which can lead to clusters of a wide age range having similar maximum luminosities \citep{Gieles_et_al_06b, Pfeffer_et_al_19a}.
As expected from their similar ages (Fig.~\ref{fig:BrightestAge}) but brighter luminosities (Fig.~\ref{fig:Brightest}), the masses of cluster candidates from observed galaxies are often larger than the median for the simulations, but still generally fall within the overall distribution for the simulated galaxies.
Similar to the simulations, the observed galaxies do not show any strong trends in the mass of the brightest cluster as a function of redshift.

In Appendix~\ref{app:BrightestMet} we also compare the metallicities of the brightest clusters.
However, due to age--metallicity degeneracies and the weak dependence of SED fitting on metallicity for young objects \citep[e.g. see][]{Adamo_et_al_23}, the current observations are not a strong test of the simulation predictions.

\subsection{Location of brightest clusters in galaxies}
\label{sec:location}

Visual inspection of images of lensed galaxies indicates that the majority of cluster candidates are coincident with the main arc of the lensed galaxies \citep[e.g.][]{Vanzella_et_al_22a, Vanzella_et_al_22b, Vanzella_et_al_23, Claeyssens_et_al_23, Adamo_et_al_24, Fujimoto_et_al_24, Messa_et_al_24b, Mowla_et_al_24}, with only the Sparkler (SMACS0723 S2) and Firework (SMACS0723 S4) galaxies displaying a large number of clumps projected off the main arc \citep[it is notable that both galaxies have significantly larger numbers of detected clumps compared to most other lensed galaxies, with both having nearly 30 clumps,][]{Mowla_et_al_22, Claeyssens_et_al_23}.
This is a general prediction for the `young cluster' scenario for GC formation in E-MOSAICS.
Comparing the galactocentric radius of the brightest clusters to the 3D half-mass radius of the galaxies ($r_{\ast,1/2}$) we find that $50$ per cent of clusters are within $\approx 0.37 r_{\ast,1/2}$, $84$ per cent of clusters (upper $1 \sigma$) are within $\approx 1.1 r_{\ast,1/2}$ and $97.5$ per cent of clusters (upper $2 \sigma$) are within $\approx 3.7 r_{\ast,1/2}$, with no strong trends as a function of redshift or observation band (F090W to F444W).
These predictions could be systematically compared in future with source plane reconstruction and galactocentric-distance measurements for the cluster candidates in lensed galaxies.

\section{Discussion}
\label{sec:discussion}

Comparisons of simulation predictions with high-redshift observations of star clusters have the potential to inform and improve current models of GC formation.
However, any limitations or biases that may affect the comparisons first need to be understood.
In this case, the offset in the median for brightest cluster luminosities (Fig.~\ref{fig:Brightest}) and masses (Fig.~\ref{fig:BrightestMass}) predicted for E-MOSAICS galaxies compared to observed lensed galaxies may have a number of different origins, such as:

\begin{itemize}

\item E-MOSAICS underestimates cluster masses: If the brightest clusters in E-MOSAICS high redshift galaxies were $\approx 0.5$~dex more massive they would agree well with the median observed cluster mass estimates (Fig.~\ref{fig:BrightestMass}). This could be motivated by uncertainties within the MOSAICS cluster formation model or EAGLE galaxy formation model, given both are calibrated to $z \approx 0$ observations.
However, alternative versions of the E-MOSAICS model which do allow for higher mass clusters (where the upper truncation in the initial cluster mass function is removed) are inconsistent with the $z=0$ properties of both young star cluster \citep[in particular the brightest cluster-galaxy star formation rate relation,][]{Pfeffer_et_al_19b} and GC populations \citep[the GC mass function truncation-galaxy mass relation,][]{Hughes_et_al_22}.
Such a change would need to apply only in high-redshift galaxies \citep[the upper cluster truncation mass is already larger in high-redshift galaxies in E-MOSAICS due to their higher gas fractions,][]{Pfeffer_et_al_19b}, and be accompanied by increased disruption of massive clusters \citep[which in E-MOSAICS occurs mainly through dynamical friction at the high-mass end of the cluster mass function,][]{P18}.

\item Underpredicted galaxy SFRs:
An offset in galaxy SFRs between the simulated and observed galaxies could potentially explain the offset in brightest cluster luminosity through the brightest cluster--SFR relation (Fig.~\ref{fig:Brightest_SFR}).
Indeed, the EAGLE model tends to slightly underpredict the SFRs of high redshift galaxies \citep{Furlong_et_al_15}.
However, for the available galaxy sample the SFRs agree reasonably well (Fig.~\ref{fig:GalSFR}).
A larger sample of observed SFRs in lensed galaxies is needed to confirm any systematic offset.
Using SED fits to calculate rest-frame $V$ band luminosities would also confirm if high-redshift galaxies follow different $V$ band brightest cluster--SFR relations at different redshifts (Fig.~\ref{fig:Brightest_SFR}).
Additionally, this could help confirm if confusion effects (see below) are elevating observed cluster brightnesses (i.e. if clusters were consistently significantly brighter than expected for the galaxies' SFRs).

\item Sampling of galaxy luminosity function: The E-MOSAICS galaxies are volume complete and therefore are naturally biased towards fainter galaxies in the luminosity selection (Fig.~\ref{fig:GalMass}). This may not be the case for observed galaxies, for which brighter galaxies are easier to detect.
Due to correlations between the brightest cluster and galaxy luminosity, sampling of the galaxy luminosity function may significantly affect the cluster luminosity distributions.
By design, the median galaxy luminosity is relatively similar between observed and simulated galaxies within the luminosity selection (Fig.~\ref{fig:GalMass}).
However, such an effect should be accounted for as the observed lensed galaxy sample becomes larger.

\item Limited simulation volume: The largest E-MOSAICS simulation volume (side length $L = 34.4 \cMpc$) is not large enough to contain progenitors of very massive galaxy clusters ($M_{200} \sim 10^{15} \Msun$), with the most massive group being $M_{200} \approx 10^{13.7} \Msun$ at $z=0$. Simulations containing such rare regions require $L \sim 1 \Gpc$ sized volumes.
However, for a given stellar mass, galaxies at higher redshifts will be found in higher mass haloes at $z=0$.
For example, in the EAGLE L100N1504 simulations \citep{S15}, galaxies with $M_\ast \approx 10^9 \Msun$ at $z=1$ will typically be progenitors of galaxies located (as central or satellites) in haloes with $M_{200} \approx 10^{12} \Msun$ at $z=0$, while similar mass galaxies at $z=6$ will typically be found in $M_{200} \approx 10^{13.5} \Msun$ haloes at $z=0$. Scaling this to higher redshifts means $z=10$ galaxies of similar mass would be expected to be progenitors of galaxies in $M_{200}(z=0) > 10^{14} \Msun$ haloes.
Potentially, the observed lensed galaxies (particularly at the highest redshifts, $z \sim 10$) may be progenitors from environments not covered by the E-MOSAICS volume. If there was a dependence of brightest cluster properties on environment at high redshift (e.g. perhaps through natal gas pressure dependence), this could bias the simulation predictions to lower masses and luminosities due to their locations in underdense regions (relative to those expected from much larger volumes). However, we note that an environmental ($M_{200}$) dependence was not found for GC mass function truncations in E-MOSAICS \citep{Hughes_et_al_22}.

\item Confusion with cluster complexes: In the Local Universe, star clusters generally do not form in isolation, but in associations termed star cluster complexes \citep[e.g.][]{Zhang_Fall_and_Whitmore_01, Bastian_et_al_05}. As observational resolution decreases, neighbouring clusters and star-forming regions become blended in imaging, leading to detected clumps with larger luminosities and sizes, as well as overestimated clump/cluster luminosity fractions \citep[e.g.][]{Cava_et_al_18, Messa_et_al_19}. Indeed, high redshift clumps (at fixed redshift) show strong trends of luminosity and size with lensing magnification \citep[e.g.][]{Mestric_et_al_22, Claeyssens_et_al_23}, indicative of such blending. Testing this effect would require higher-still resolution imaging (e.g. with future extremely large telescopes) or limiting comparisons to lensed galaxies with the very highest resolutions, which would severely limit the galaxy sample.

\item Lens stretching: Related to the previous issue, the brightness of lensed star clusters may be overestimated due to lens stretching (magnification/resolution is higher in the tangential direction than the radial direction). For example, \citet{Vanzella_et_al_19} found that fluxes may be overestimated by a factor $\approx 1.3$--$1.5$ ($0.3$--$0.45 \Mag$ brighter) when modelling the lensing of a nearby dwarf galaxy (NGC 1705) with a young, massive star cluster. The exact amount of overestimation will naturally depend on factors such as resolution, the ratio of tangential and radial magnifications and particular orientation of the system, so will not be a fixed offset in all galaxies and needs to be estimated separately in each case.

\item Cluster detection limits: In half of the observed galaxies in Fig.~\ref{fig:Brightest} within the luminosity selection limits (Fig.~\ref{fig:GalMass}), the faintest detected star cluster candidate is similar to or brighter than the median prediction from E-MOSAICS. If the faintest detected clusters are similar to detection limits, this could imply only around half of lensed galaxies have detectable clusters and represent mainly the upper half of the distribution, those with the very brightest star clusters.
Many lensed galaxies are indeed excluded from the sample due to not containing compact clumps \citep[e.g. from][]{Claeyssens_et_al_23, Messa_et_al_24}, though interpretation is complicated by the differing magnifications and resolutions between different galaxies.
Detection effects could be further tested by determining the distribution of cluster luminosity detection limits in lensed galaxies at fixed spatial resolution, to take into account varying magnifications and redshifts.
Fainter clusters could also be detected with deeper imaging (where magnification/resolution is sufficient).

\item Uncertainties in stellar population models: We use simple stellar populations from the FSPS model \citep{Conroy_Gunn_and_White_09, Conroy_and_Gunn_10} to calculate cluster luminosities as it includes \textit{JWST} filters. The default model with Padova isochrones does not include effects such as binary star evolution or stellar rotation, which are more important at lower metallicities and could increase the brightness of stellar populations in the UV \citep{Levesque_et_al_12, Stanway_et_al_16, Eldridge_et_al_17}.
Similarly, in very young populations, ionised gas can be of similar importance as the stellar light in UV and optical bands \citep{Reines_et_al_10}, thus uncertainties in its modelling (covering fraction, etc.) could have large effects in the stellar population models.
A lack of (or underestimate of) such effects in stellar population modelling may simultaneously affect both the simulations (lower than expected luminosities) and observations (higher than expected masses).
We do not test the effect of using BPASS isochrones \citep{Eldridge_et_al_17} in FSPS as they have a fixed Salpeter IMF, leading to higher $M/L$ at all ages relative to a Chabrier IMF, even with binary star evolution. However, relative to a Salpeter IMF with Padova isochrones, the BPASS isochrones lead to luminosities that are $\approx20$--$60$ per cent brighter between ages of $1$--$10^3 \Myr$. This luminosity increase is potentially enough to account for the offset between observations and simulation predictions in both luminosity and mass, although, as we discuss next, changes in mass-to-light ratios can be compensated by changes in galaxy mass selection (for fixed luminosity selection).
One effect that might apply predominantly to star clusters is an increase in UV luminosities due to helium-enhanced stars in proto-GCs \citep{Katz_et_al_24}, but more work is needed to understand the extent of the effect for stellar populations at different ages and metallicities, and in different wavebands.

\item Uncertainties in stellar IMF: Some works suggest the stellar IMF may be top-heavy at high redshifts \citep[e.g.][]{Cameron_et_al_24, Mowla_et_al_24}. As with effects like binary star evolution, a top-heavy IMF will make young stellar populations more luminous for a given mass. We tested this with FSPS by using a Kroupa IMF with a flatter high mass power-law index of $\alpha = -1.5$ (noting that this is no longer consistent with the Chabrier IMF used in the EAGLE model, which would require also recalibrating the stellar feedback model, e.g. as in \citealt{Barber_Crain_and_Schaye_18}). We find that a top-heavy IMF does not significantly change the predictions for brightest cluster luminosities, as the lower mass-to-light ratios leads to selection of lower-mass galaxies (in this case by a factor $\approx 2$) with lower-mass clusters.
Instead, the main effects of a top-heavy IMF are that the brightest clusters become significantly younger (median ages $<10 \Myr$ at all redshifts $z \geq 1$ for F150W, $<20 \Myr$ for F444W) and lower mass (median masses $\sim 10^5 \Msun$).

\item Dust/extinction uncertainties:
In most cases \citep[except for, e.g., the Cosmic Grapes, where dust maps are derived from ALMA observations,][]{Fujimoto_et_al_24}, extinction values are derived from SED fitting, and are therefore degenerate with other properties. Systematic offsets in both clusters and the galaxies would be unlikely to explain the brightest cluster luminosity offsets, for the same reasoning as IMF variations (it would also change galaxy selection).
However, there is some evidence from local galaxies that higher-mass young star clusters clear their surroundings of absorbing gas/dust earlier than lower-mass clusters \citep{McQuaid_et_al_24}. If such a process was at work in young clusters in the high-redshift Universe and also extended to the field star regime, then, for a given luminosity, galaxies would be more massive and therefore expected to have higher mass clusters. In practice, this would require extinction to be underestimated for the whole galaxy, or overestimated for the star cluster candidates, given that extinction is generally similar for both in the lensed galaxies in Section~\ref{sec:observations}.
Alternatively, simulations with explicit models for dust, combined with radiative transfer calculations, would enable direct comparison of observed luminosities without extinction corrections.

\item Nuclear star clusters: Currently, high-redshift observations do not distinguish between normal star clusters and nuclear star clusters. Nuclear clusters are often the brightest and most massive star cluster in a galaxy and thought to form through either mergers of inspiralling star clusters or central star formation \citep[see][for a review]{Neumayer_et_al_20}, but are not modelled in E-MOSAICS. In some cases the brightest cluster candidates appear coincident with the centre of the host galaxies \citep[e.g.][]{Fujimoto_et_al_24, Messa_et_al_24b}, indicating they could be nuclear star cluster candidates and not directly comparable with globular cluster progenitors.

\item Ultra-compact dwarf galaxies:
Related to nuclear star clusters are ultra-compact dwarf galaxies (UCDs), the most massive of which are thought to form as nuclear clusters before their host galaxies are tidally disrupted during galaxy mergers \citep[e.g.][]{Bekki_et_al_03, Drinkwater_et_al_03, Brodie_et_al_11}. 
Being typically more massive than GCs, UCDs can lead to a high-mass tail in the GC mass function \citep{Pfeffer_et_al_16} and could potentially lead to brighter than expected clusters in high-redshift galaxies.
The very young ages of most high-redshift clusters (Fig.~\ref{fig:BrightestAge}) makes a UCD explanation unlikely in most cases, as the timescales are too short for host galaxy disruption to occur.
However, a large fraction of UCD formation is expected to occur between redshifts $1$--$3$ \citep{Pfeffer_et_al_14, Mayes_et_al_21} and could contribute in more luminous galaxies in that epoch (e.g. SMACS0723 S4).

\end{itemize}

At present, it is not possible to determine what is the most important effect in explaining the offset between simulated and observed cluster luminosities and masses, but this discussion may guide future tests.
Potentially, some combination of effects may explain the offset (for example, lens stretching may already explain $\approx 0.3$--$0.5 \Mag$ difference, or around one third of the offset).
Given the current sample of observed high-redshift galaxies with compact clusters is relatively small, statistical comparisons will improve as more observations are taken \citep[e.g.][]{Claeyssens_et_al_24, Naidu_et_al_24} and any selection effects can be better understood.

Our test varying the stellar IMF also shows that global changes to stellar populations may not change the predictions for the luminosities of the brightest clusters, as it also changes the galaxy selection.
In this case, to change the brightest cluster predictions, modifications to the stellar populations would need to only (or predominantly) apply to star clusters but not field stars.
For example, star clusters could have higher binary or rotating star fractions or a more top-heavy IMF relative to field stars.
However, we caution that modifications to the stellar IMF would also modify the host galaxy properties, and a complete test requires recalibrating the stellar feedback model \citep[e.g.][]{Barber_Crain_and_Schaye_18}.

\section{Summary}
\label{sec:summary}

In this work we have compared the properties of star cluster candidates in lensed, high redshift galaxies from \textit{JWST} and \textit{HST} observations with predictions of star cluster properties from the E-MOSAICS simulations.
Such high-redshift star clusters are thought to be analogues of today's old GCs observed soon after formation, enabling tests of GC formation models.
We focus on the properties of the brightest cluster in each galaxy, so that comparisons can be made for observed galaxies with few detected compact sources.

We find that the luminosities (Fig.~\ref{fig:Brightest}), ages (Fig.~\ref{fig:BrightestAge}) and masses (Fig.~\ref{fig:BrightestMass}) of the brightest clusters in observed lensed galaxies are consistent with the E-MOSAICS predictions.
We predict that the brightest cluster--galaxy SFR relation evolves with redshift (Fig.~\ref{fig:Brightest_SFR}) such that, at a given SFR, clusters are brighter at higher redshifts, which could be tested with larger lensed galaxy samples.
For each of these properties, the observed star cluster candidates fall within the distribution predicted by the simulations.
In particular, the brightest cluster ages agree very well between observed and simulated galaxies.
Assuming the clusters survive to the present day, the ages at $z=0$ for the clusters in the sample of lensed galaxies span the range $\approx 9$--$13.5 \Gyr$ (Fig.~\ref{fig:BrightestFormTime}), indicating that GC formation may occur across a wide range of redshifts rather than at a specific epoch.
This provides further evidence that standard young star cluster formation mechanisms found in low-redshift galaxies, operating in the more extreme star formation conditions at high redshift, may explain the GCs observed in the present day.

However, the observed brightest cluster candidates tend to be brighter in the NIRCam bands than the median predicted from E-MOSAICS by around $1$--$1.5 \Mag$ (Fig.~\ref{fig:Brightest}).
This is similarly reflected in the estimated masses of the brightest clusters in F150W, with around $0.5$~dex offset in mass (Fig.~\ref{fig:BrightestMass}, right panel).
As discussed in Section~\ref{sec:discussion}, such a difference could be explained by many effects, including underestimated cluster masses or galaxy SFRs in the simulations, nuclear star clusters, uncertainties in stellar population modelling, observational resolution limitations and selection effects of observed galaxies.
We also tested the effect of a top-heavy IMF for simulation luminosity estimates, but found the brighter stellar populations were compensated by selection of lower mass galaxies and star clusters, such that the brightest cluster luminosities were similar.

Clearly, before high-redshift observations of star clusters can motivate modifications to, or confirm the accuracy of, present GC formation models, systematic effects or biases in the comparisons must first be understood.
Future studies should work to understand such limitations to enable stronger tests of GC formation models.

\section*{Acknowledgements}

We thank Matteo Messa and Seiji Fujimoto for providing luminosity catalogues for MACSJ0416 and Cosmic Grapes.
We thank the referee for a helpful report which improved the paper.
This work was supported by the Australian government through the Australian Research Council's Discovery Projects funding scheme (DP220101863) and the Australian Research Council Centre of Excellence for All Sky Astrophysics in 3 Dimensions (ASTRO 3D), through project number CE170100013.
AJR was supported by National Science Foundation grant AST-2308390.  Support for Program number HST-GO-15235 was provided through a grant from the STScI under NASA contract NAS5-26555.
JMDK gratefully acknowledges funding from the European Research Council (ERC) under the European Union's Horizon 2020 research and innovation programme via the ERC Starting Grant MUSTANG (grant agreement number 714907). COOL Research DAO is a Decentralized Autonomous Organization supporting research in astrophysics aimed at uncovering our cosmic origins.
MC gratefully acknowledges funding from the DFG through an Emmy Noether Research Group (grant number CH2137/1-1).
This work used the DiRAC Data Centric system at Durham University, operated by the Institute for Computational Cosmology on behalf of the STFC DiRAC HPC Facility (\url{www.dirac.ac.uk}). This equipment was funded by BIS National E-infrastructure capital grant ST/K00042X/1, STFC capital grants ST/H008519/1 and ST/K00087X/1, STFC DiRAC Operations grant ST/K003267/1 and Durham University. DiRAC is part of the National E-Infrastructure.
The work also made use of high performance computing facilities at Liverpool John Moores University, partly funded by the Royal Society and LJMU's Faculty of Engineering and Technology.
Some of the data products presented herein were retrieved from the Dawn JWST Archive (DJA). DJA is an initiative of the Cosmic Dawn Center (DAWN), which is funded by the Danish National Research Foundation under grant DNRF140.

\section*{Data Availability}

The data underlying this article will be shared on reasonable request to the corresponding author.


\bibliographystyle{mnras}
\bibliography{emosaics}



\appendix

\section{Lensed galaxy properties}
\label{app:Table}

Table~\ref{tab:observations} summarises the properties of observed lensed galaxies and their star cluster candidates from Section~\ref{sec:observations}.

\begin{landscape}

\setlength{\tabcolsep}{5pt} 
\renewcommand{\arraystretch}{1.5} 
\begin{table}

\caption{Summary of observed lensed galaxies described in Section~\ref{sec:observations}: 
(1) Galaxy name/ID; 
(2) redshift; 
(3) galaxy observed-frame absolute magnitude in NIRCAM F150W (corrected for redshift, magnification, dust extinction); 
(4) galaxy stellar mass;
(5) galaxy star formation rate;
(6) number of star cluster candidates;
(7) observed-frame absolute magnitudes of the brightest cluster in F150W;
(8) age of the brightest cluster in F150W;
(9) stellar mass of the brightest cluster in F150W;
(10) observed-frame absolute magnitude of the brightest cluster in F444W;
(11) age of the brightest cluster in F444W;
(12) stellar mass of the brightest cluster in F444W;
(13) total cluster luminosity fraction in F150W, $f_\mathrm{cl,F150W} = (\sum L_\mathrm{cl,F150W}) / L_\mathrm{gal,F150W}$;
(14) total cluster luminosity fraction in F444W, $f_\mathrm{cl,F444} = (\sum L_\mathrm{cl,F444W}) / L_\mathrm{gal,F444W}$;
(15) total cluster stellar mass fraction, $f_{M_\ast} = (\sum M_\mathrm{\ast,cl}) / M_\mathrm{\ast,gal}$.
${}^\mathrm{a}$ For A2744 S3 NIRISS F200W band is used instead of NIRCAM F150W. ${}^\mathrm{b}$ For the Sunburst arc we use the brightest cluster in \textit{HST} F555W. ${}^\mathrm{b}$ For RCS0224 and MACS0940 arcs we use the brightest clusters in \textit{HST} F814W.}
\label{tab:observations}
\begin{tabular}{@{}lcccccccccccccc}
\hline
Galaxy & $z$ & $M_\mathrm{gal,F150W}$ & $M_\mathrm{\ast,gal}$ & SFR & $N_\mathrm{clust}$ & $M_\mathrm{br,F150W}$ & Age$_\mathrm{br,F150W}$ & $M_\mathrm{\ast,br,F150W}$ & $M_\mathrm{br,F444W}$ & Age$_\mathrm{br,F444W}$ & $M_\mathrm{\ast,br,F444W}$ & $f_\mathrm{cl,F150W}$ & $f_\mathrm{cl,F444W}$ & $f_{M_\ast}$ \\
 & & [$\Mag$] & [$10^8 \Msun$] & [$\rmn{M}_{\sun} \pyr$] & & [$\Mag$] & [Myr] & [$10^6 \Msun$] & [$\Mag$] & [Myr] & [$10^6 \Msun$] & & & \\
(1) & (2) & (3) & (4) & (5) & (6) & (7) & (8) & (9) & (10) & (11) & (12) & (13) & (14) & (15) \\
\hline
Cosmic Gems & $10.2$ & $-18.39$ & $0.24$--$0.56$ & $0.33^{+0.03}_{-0.09}$ & $5$ & $-16.14$ & $9.2^{+13.5}_{-3.2}$ & $2.45^{+5.20}_{-1.56}$ & $-15.45$ & $9.2^{+13.5}_{-3.2}$ & $2.45^{+5.20}_{-1.56}$ & $0.305$ & $0.365$ & $0.209^{+0.542}_{-0.141}$ \\
Firefly & $8.304$ & $-19.94$ & $0.063^{+0.239}_{-0.028}$ & $0.63^{+2.53}_{-0.00}$ & $10$ & $-17.39$ & $121^{+62.5}_{-121}$ & $0.631^{+3.350}_{-0.315}$ & $-17.65$ & $84.6^{+98.8}_{-84.6}$ & $0.501^{+2.011}_{-0.250}$ & $0.442$ & $0.400$ & $0.640^{+2.564}_{-0.355}$ \\
MACSJ0416 D1 & $6.144$ & $-17.47$ & $0.302^{+0.032}_{-0.057}$ & $0.34^{+0.08}_{-0.03}$ & $1$ & $-15.96$ & $12^{+0}_{-1}$ & $5.5^{+0.2}_{-0.3}$ & & & & $0.249$ & & $0.182^{+0.007}_{-0.010}$ \\
MACSJ0416 T1 & $6.145$ & $-16.12$ & $0.010^{+0.001}_{-0.000}$ & $0.82^{+0.02}_{-0.19}$ & $1$ & $-14.63$ & $2^{+6}_{-0}$ & $0.2^{+0.8}_{-0.0}$ & & & & $0.253$ & & $0.200^{+0.800}_{-0.000}$ \\
MACSJ0416 UT1 & $6.145$ & $-14.73$ & $0.049^{+0.056}_{-0.006}$ & $0.01^{+0.01}_{-0.01}$ & $1$ & $-14.48$ & $14^{+57}_{-0}$ & $1.4^{+6.4}_{-0.0}$ & & & & $0.794$ & & $0.286^{+1.306}_{-0.000}$ \\
Cosmic Grapes & $6.072$ & $-19.80$ & $4.5^{+2.7}_{-1.1}$ & $2.6^{+1.7}_{-1.5}$ & $4$ & $-18.78$ & $6.9^{+2.7}_{-2.7}$ & $81.9^{+3.3}_{-3.3}$ & & & & $0.487$ & & $0.333^{+0.017}_{-0.017}$ \\
Sunrise & $\approx 6$ & $-19.50$ & $3$--$22$ & $3$--$10$ & $3$ & $-17.03$ & $4^{+2}_{-3}$ & $3.3^{+3.2}_{-0.8}$ & $-16.79$ & $4^{+2}_{-3}$ & $3.3^{+3.2}_{-0.8}$ & $0.173$ & $0.140$ & $0.011^{+0.053}_{-0.008}$ \\
A2744 S3$^\mathrm{a}$ & $3.981$ & $-19.87$ & $1.648^{+0.423}_{-1.255}$ & $1.47^{+0.85}_{-0.25}$ & $3$ & $-16.74$ & $79^{+21}_{-74}$ & $4.1^{+0.9}_{-3.3}$ & & & & $0.087$ & & $0.035^{+0.020}_{-0.024}$ \\
Sparkler & $1.378$ & $-19.39$ & $5$--$10$ & & $12$ & $-14.52$ & $200^{+101}_{-194}$ & $10.00^{+5.85}_{-6.84}$ & $-14.77$ & $200^{+101}_{-194}$ & $10.00^{+5.85}_{-6.84}$ & $0.063$ & $0.104$ & $0.138^{+0.093}_{-0.052}$ \\
SMACS0723 S4 & $2.31$ & $-21.23$ & & & $10$ & $-15.25$ & $2^{+48}_{-0}$ & $0.63^{+0.37}_{-0.53}$ & $-14.5$ & $1007^{+1007}_{-300}$ & $12.59^{+3.26}_{-2.59}$ & $0.009$ & $0.003$ & \\
SMACS0723 S1 & $1.449$ & $-20.39$ & & & $1$ & $-16.55$ & $20^{+10}_{-9}$ & $15.85^{+4.10}_{-3.26}$ & $-17.41$ & $20^{+10}_{-9}$ & $15.85^{+4.10}_{-3.26}$ & $0.029$ & $0.058$ & \\
SMACS0723 S5 & $1.425$ & $-20.00$ & & & $3$ & $-12.31$ & $80^{+20}_{-10}$ & $0.63^{+0.16}_{-0.13}$ & $-12.75$ & $10^{+5}_{-7}$ & $0.063^{+0.063}_{-0.023}$ & $0.002$ & $0.003$ & \\
SMACS0723 S7 & $5.17$ & & & & $3$ & $-14.75$ & $15^{+5}_{-8}$ & $1.59^{+0.41}_{-0.95}$ & $-14.06$ & $15^{+5}_{-8}$ & $1.59^{+0.41}_{-0.95}$ & & & \\
SMACS0723 I8 & $2.12$ & & & & $1$ & $-13.46$ & $20^{+10}_{-16}$ & $0.79^{+0.79}_{-0.16}$ & $-14.51$ & $20^{+10}_{-16}$ & $0.79^{+0.79}_{-0.16}$ & & & \\
SMACS0723 S3 & $1.991$ & & & & $2$ & $-19.17$ & $2^{+1}_{-1}$ & $19.95^{+5.17}_{-0.00}$ & $-17.52$ & $20^{+10}_{-6}$ & $19.95^{+0.00}_{-4.10}$ & & & \\
\hline
Sunburst$^\mathrm{b}$ & $2.37$ & & $10$ & $9.95^{+13.42}_{-3.16}$ & $9$ & & $3^{+0}_{-0}$ & $11.7^{+11.7}_{-0.0}$ & & & & & & $0.030^{+0.030}_{-0.000}$ \\
RCS0224$^\mathrm{c}$ & $4.88$ & & & & $2$ & & $10.5^{+9.5}_{-9.5}$ & $31.62^{+18.50}_{-26.61}$ & & & & & & \\
MACS0940$^\mathrm{c}$ & $4.03$ & & & & $3$ & & $25.5^{+14.5}_{-14.5}$ & $31.62^{+284.60}_{-6.50}$ & & & & & & \\
\hline
\end{tabular}
\end{table}
\end{landscape}

\section{Brightest cluster metallicities}
\label{app:BrightestMet}

\begin{figure*}
  \includegraphics[width=0.495\textwidth]{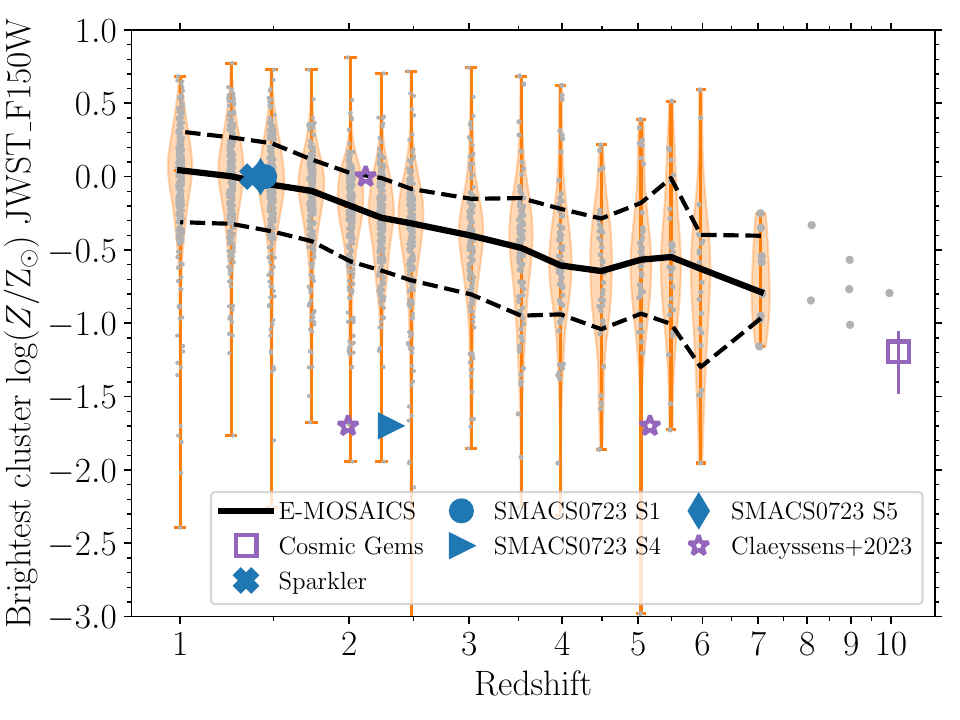}
  \includegraphics[width=0.495\textwidth]{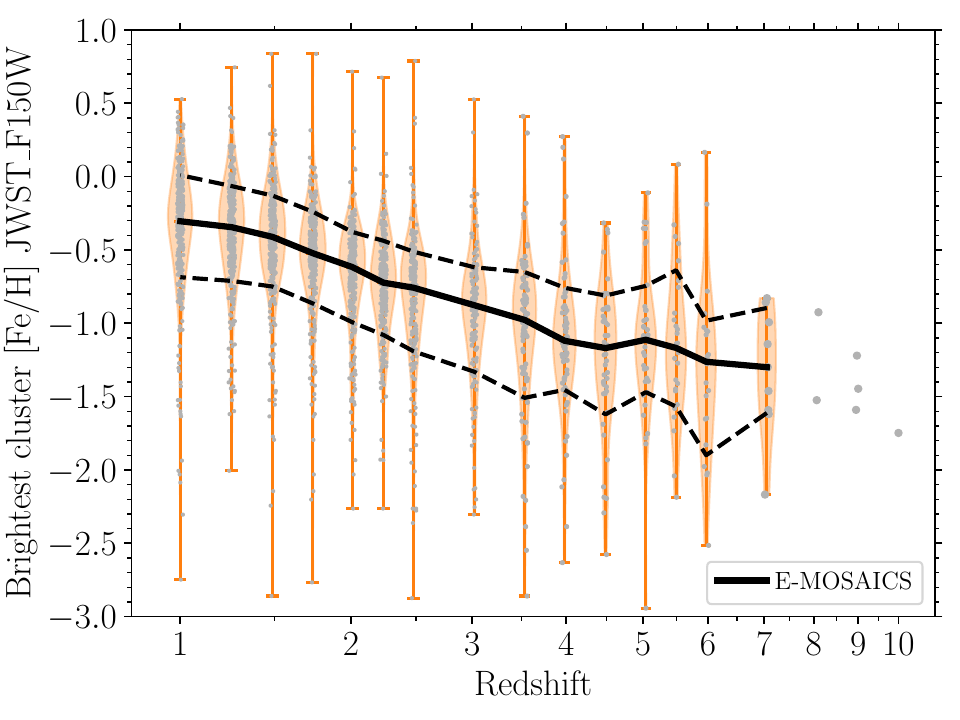}
  \caption{Metallicities (left: $\LogZ$; right: $\FeH$) of the brightest clusters in F150W. Point and line styles are as in Fig.~\ref{fig:GalMass}.}
  \label{fig:BrightestMet}
\end{figure*}

Fig.~\ref{fig:BrightestMet} compares the metallicities of the brightest cluster in F150W in each simulated galaxy as a function of redshift.
We do not find any significant variations in the median metallicities with observation band.
The typical metallicities of the brightest clusters increase with decreasing redshift.
This is expected from the increasing masses of the galaxies (Fig.~\ref{fig:GalMass}) and the galaxy mass--metallicity relation for EAGLE galaxies \citep{S15}.

For reference, we also include the metallicity estimates for cluster candidates in lensed galaxies, where available.
Other than Cosmic Gems, all other lensed galaxies shown in the figure are from \citet{Claeyssens_et_al_23}, though these values are highly uncertain as only four metallicities were tested ($\LogZ = [0, -0.4, -0.7, -1.7]$) and the SED fits often only weakly depend on metallicity (see their figure F1).
Other works have also analysed the Sparkler galaxy, including \citet[finding metallicities in the range $\LogZ \approx -1$ to $0.2$]{Mowla_et_al_22} and \citet[finding metallicities in the range $\FeH \approx -2$ to $-0.2$]{Adamo_et_al_23}.
These values are broadly in agreement with those found in the simulated galaxies at $z \approx 1.5$.
However, age--metallicity degeneracies generally mean the metallicities from SED fitting are very uncertain, particularly for young ($<1\Gyr$) objects \citep[e.g. see figure 3 in][]{Adamo_et_al_23}.
These comparisons could be improved in future with spectroscopic metallicities.


\bsp	
\label{lastpage}
\end{document}